\documentclass[paper=letter, fontsize=12pt]{article} 
\RequirePackage[T1]{fontenc}
\RequirePackage{amsthm,amsmath,amsfonts}
\usepackage{longtable}
\usepackage{graphicx}

\usepackage{latexsym}
\usepackage{multirow}

\RequirePackage[numbers]{natbib}
\PassOptionsToPackage{hyphens}{url}\RequirePackage[colorlinks,citecolor=blue,urlcolor=blue]{hyperref}

\allowdisplaybreaks[4]

\usepackage{times}
\usepackage{keyval}
\usepackage{calc}

\usepackage[letterpaper]{geometry}
\renewcommand{\baselinestretch}{1.1}
\numberwithin{equation}{section}
\theoremstyle{plain}

\usepackage{fancyhdr}
\usepackage{adjustbox}
\usepackage{float}
\usepackage{enumitem} 
\newcommand{\supervisor}{Cody Hyndman} %
\newcommand{\coauthorT}{Traian A Pirvu} %
\newcommand{\coauthorP}{Petar Jevti\'{c}} %
\newcommand{\papertitle}{Optimal annuitization post-retirement with labor income} %
\makeatletter %
\@addtoreset{equation}{section}
\makeatother  %
\theoremstyle{plain}
\newtheorem{theorem}{Theorem}[section]%

\newtheorem{defn}[theorem]{Definition} %

\newtheorem{lemma}[theorem]{Lemma}
\newtheorem{corollary}[theorem]{Corollary}
\newtheorem{remark}[theorem]{Remark}
\newtheorem{variational inequality}[theorem]{Variational inequality}

\DeclareMathAlphabet{\pazocal}{OMS}{zplm}{m}{n}
\DeclareSymbolFontAlphabet{\mathrm}    {operators}
\DeclareSymbolFontAlphabet{\mathnormal}{letters}
\DeclareSymbolFontAlphabet{\mathcal}   {symbols}
\DeclareMathAlphabet{\mathbf}{OT1}{cmr}{bx}{n}
\DeclareMathAlphabet{\mathsf}{OT1}{cmss}{m}{n}
\DeclareMathAlphabet{\mathit}{OT1}{cm}{m}{it}
\DeclareMathAlphabet{\mathpzc}{OT1}{pzc}{m}{it}
\DeclareGraphicsExtensions{.eps}
\begin{document}
	\rhead{\textit{Jan.\ 29, 2022}}
	\lhead{\textit{Gao, Hyndman, Pirvu, Jevti\'{c}}}
	\chead{\textit{Optimal annuitization with labor income}}

	\title{\papertitle}
	
	\author{
		Xiang Gao\footnote{ 
			Department of Mathematics and Statistics, 
			Concordia University, 
			Montr\'eal, Qu\'ebec.
			emails: cody.hyndman@concordia.ca; xiang.gao@concordia.ca
		}
		, 
		\supervisor\footnotemark[1]\ \footnote{Corresponding Author}
		,
		\coauthorT\footnote{Department of Mathematics \& Statistics, McMaster University, Hamilton, Ontario. email: tpirvu@math.mcmaster.ca}
		\ and
		\coauthorP\footnote{School of Mathematical and Statistical Sciences, Arizona State University, Tempe, Arizona. email: petar.jevtic@asu.edu}
	}
	
	\date{January 29, 2022}
	
	\maketitle

	\abstract{

	  Evidence shows that the labor participation rate of retirement age cohorts is non-negligible, and it is a widespread phenomenon globally. %
          In the United States, the labor force participation rate for workers age 75 and older is projected to be over 10 percent by 2026 as reported by
the Bureau of Labor Statistics.
          The prevalence of post-retirement work changes existing considerations of optimal annuitization, a research question further complicated by novel factors such as post-retirement labor rates, wage rates, and capacity or willingness to work. To our knowledge, this poses a practical and theoretical problem not previously investigated in actuarial literature. In this paper, we study the problem of post-retirement annuitization with extra labor income in the framework of stochastic control, optimal stopping, and expected utility maximization. The utility functions are of the Cobb-Douglas type. The martingale methodology and duality techniques are employed to obtain closed-form solutions for the dual and primal problems. The effect of labor income is investigated by exploiting the explicit solutions and Monte-Carlo simulation. The latter reveals that the optimal annuitization time is strongly linear with respect to the initial wealth, with or without labor income.  When it comes to optimal annuitization, we find that the wage and labor rates may play opposite roles. However, their impact is mediated by the leverage ratio. 
	}
	
	\vspace{5mm}
	
	\noindent
	\textbf{Keywords:}
	Stochastic control; Optimal stopping; Post-retirement annuitization; Cobb–Douglas utility; Labor income; Martingale methods.

	\vspace{5mm}
	\noindent
	\textbf{Mathematics Subject Classification (2020):}
	Primary: 91G05, 93E20; Secondary: 60G40, 91B16

	\renewcommand{\baselinestretch}{1.5}

	\section{Introduction}
	The labor participation rates of retirement age cohorts is non-negligible and varies geographically. In 2020, the OECD \cite{oecd1} reported that
        the labor participation rate of age 65 and over ranges from 2.8\% in Luxembourg, 7.4\% in Germany, 10.7\% in the United Kingdom, 13.8\% in Canada,
        35.5\% in Korea, and  41.7\% in Indonesia. Statistics Canada \cite{stat2} reports that the total number of employed persons in Canada who are
        65 years and older increased from 212.8 thousand in 2001 to 874.2 thousand in 2021. Many demographic, sociological, economic, and other factors
        contribute to old-age employment, and post-retirement, or bridge, employment (see \cite{pr1,pr2,pr3,pr4,pr5}).
         According to the U.S.\ Bureau of Labor Statistics \cite{bls1}, the labor force participation rate for workers age 75 and older is projected to be over 10 percent by 2026. In a 2021 Gallup survey \cite{gal1} of U.S.\ retirees and non-retirees, among non-retirees 21 percent expect that their main source of income will be from part-time work. Another recent report \cite{stat1} showed that in the U.S.\ as of 2020, 37 percent of survey respondents expected the primary source of retirement income would come from continuing to work.  Therefore, the evidence suggests that the phenomenon of retirement-age employment and post-retirement work will continue unabated and this factor should be included in future actuarial research on retirement planning.
	
	In this paper we consider the utility maximization problem in a defined contribution scheme where retirees balance consumption and investment with a limited capacity to gain extra labor income after retirement until their wealth reachs the target single payment annuity premium. Utility maximization problems using stochastic control and optimal stopping goes back at least to the seminal article of \citet{merton1971optimum} and has been studied extensively in past decades. For example, \citet{pliska1986stochastic} studies its application in optimal trading, \citet{karatzas1987optimal} studies the optimal portfolio and consumption decision in very explicit feedback form and \citet{cox1989optimal} study the consumption-portfolio problem when asset prices follow a diffusion process. The current literature dealing with financial risk of defined contribution pension schemes in a stochastic framework is quite rich, for instance see \citet{cairns2006stochastic}, \citet{gao2008stochastic} and \citet{gerrard2012choosing}. The benefit of including labor income is to help interpret the changes to the optimal solution more realistically when the labor effort and wage rate impact retirement decisions.
	
	Related work by \citet{karatzas2000utility} demonstrates the application of the martingale method and Legendre transforms in solving mixed optimal stopping/control problems for utility maximization. \citet{karatzas2000utility} introduced the shadow process and the corresponding budget constraint which is applied to relax the control terms from the objective function and obtain a deterministic problem. \citet{he1993labor} studied optimal investment with borrowing constraints. \citet{bodie1992labor} studied an optimal problem with flexibility in labor supply and demonstrated the dependence of agent's risk tolerance on flexibility of labor supply. \citet{bodie2004optimal} studied the optimal consumption and investment problem in a context of retirement which has a fixed time of retirement rather than an optimal time of retirement. \citet{farhi2007saving} studied the binomial choice of leisure for a Cobb-Douglas utility function, which is more generally used in measuring labor and capital inputs in economic production and is a special case of the constant elasticity of substitution (CES) utility function. \citet{choi2008optimal} studied a similar problem as \citet{farhi2007saving} in optimal consumption-leisure, portfolio, and retirement choice of an infinitely lived investor whose instantaneous utility is given by a CES utility function of consumption and leisure. \citet{choi2008optimal} uses combination of portfolio and consumption-leisure choice and an optimal stopping time for retirement problem, in which the investor derives utility from adjusting between consumption and leisure, and also has an option for fully retirement from labor. \citet{choi2008optimal} provided a solution to the free boundary problem using the martingale approach, however, lacking a solid proof of the existence and uniqueness. \citet{koo2013optimal} studied, in the Cobb-Douglas utility functions paradigm, the voluntary retirement problem with optimal investment, consumption, leisure, and retirement time, given a fixed constant leisure post-retirement.

	Let us present the contributions of our paper. We provide the methodology to consider the problem of expected utility maximization of post-retirement annuitization, labor, consumption, and investment with Cobb-Douglas type utilities. We solve this problem in closed form and the effect of different model parameters is investigated through numerical experiments. We extend the model in \citet{gerrard2012choosing} to allow for labor income, and the martingale method of \citet{choi2008optimal} is adapted to our framework with rigorous proofs that are lacking in most papers. A numerical analysis of the utility surface with varying labor rate and wage rate is carried out by exploiting closed form solutions, that reveals the optimal attainable annuity premium is concave in labor rate and convex in wage rate. Monte-Carlo simulation shows that the optimal annuitization time is strongly linear with respect to the initial wealth, in cases with and without labor income. The simulation results also reveal a critical wealth level below which the extra labor income decreases the annuitization time, and above which the extra labor income postpones the annuitization time in exchange for higher annuities. Analysis of different scenarios where we vary the leverage ratio, consumption preferences, labor preferences, and other key parameters illustrate that different pension schemes could lead to very differently optimal investment strategies and annuitization times that can be used to inform retirement and post-retirement decisions.

	\section{Economic background}
	The classical consumption-investment problem is described by an expectation of the accumulative running profit $f(\cdot)$ plus the terminal value $g(\cdot)$
	\begin{equation*}
		J\left(x|c, \tau\right) = \mathbb{E}\left[\int_0^\tau f\left(t, c_t, X^{x}_t\right) dt+ g(\tau, X^{x}_\tau)\right],
	\end{equation*}
	where the portfolio value $X^{x}_t$ is a stochastic process with initial value $x$ which is controlled by a term $c_t$ standing for the output from the system
	\begin{equation*}
		dX^{x}_t = \mu\left(t, c_t, X^{x}_t\right) dt + \sigma\left(t, c_t, X^{x}_t\right) dW_t.
	\end{equation*}
	The goal is to choose a pair of control and a stopping time $\left(c^*, \tau^*\right)$  that maximize the objective function
	\begin{equation*}
		\left(c^*, \tau^*\right) = \arg \sup_{c,\tau \in \mathbf{U}} J\left(x|c, \tau\right),
	\end{equation*}
	where all the available pairs of $\left(c, \tau\right)$ are from a space of feasible set $\mathbf{U}$. This can be generally defined as
	\begin{equation*}
		\mathbf{U} = \left\{\left(c, \tau\right)\left|\text{such that $X_t$ and $J\left(x|c, \tau\right)$ are well defined}\right. \right\},
	\end{equation*}
	which will be made more precise. Theoretically a unique solution exists if the function $f$ and $g$ are Lipschitz continuous and the process $X_t$ has bounded quadratic variation for all $\left(c, \tau\right) \in \mathbf{U}$. 
	
	The output control $c_t$ stands for the consumption rate which is the portion of the agent's total wealth consumed during one unit time. Except for the output control $c_t$, the consumption, we consider the input control $b_t$, standing for the labor income, and the input control $\pi_t$, standing for the amount of wealth invested in a risky asset. We introduce a parameter, the maximum labor rate $\bar b \in[0,1]$, to cap the labor rate from above $0 \leq b_t \leq \bar b$. Suppose the agent receives a lump sum of size $x$ at retirement when $t=0$. Up until the time of annuitization, the agent can choose to keep working after retirement with labor rate $b_t$, consume at the rate of $c_t$ and invest some amount of his or her wealth $\pi_t$ in the financial market. We assume that the remaining lifetime of the agent, $T_D$, is independent of the financial market and exponentially distributed with force of mortality $\delta$. In defined contribution pension schemes, the agent has the possibility to defer the annuitization of his wealth. The objective function of the agent before annuitization consists of the utility from consumption, dis-utility from labor, and the final utility from annuitization. The optimization problem is to maximize the agent's total expected utility from running  consumption,  extra labor income, investment gains, and the final annuity. 
	
	For simplicity, we suppose that the wage rate $w_t$ and the annuity scheme $k$, defined as a portion of the final wealth, are fixed at the agent's retirement. The financial market consists of a risk-less asset and a risky asset. The risk-less asset pays interest at fixed rate $r$ which can be the money market account or a locked in retirement account. The risky asset can be a portfolio or funds from a market which evolves as a geometric Brownian motion with mean $\mu$ and volatility $\sigma$,
	\begin{equation*}
		\frac{dS_t}{S_t}=\mu dt+\sigma dB_t.
	\end{equation*}
	Here, given the probability space $(\Omega,\mathcal{F},\mathbb{P})$ and a terminal time $T>0$, $B_t$ is a standard Brownian motion in $\mathbb{R}^d$ under the probability $\mathbb{P}$.
	
	Given an initial endowment $x \geq 0$, an income stream from labor work $wb_t$, a consumption rate $c_t$, an investment policy $\pi_t$, the remaining wealth stays in a bank account. The agent's total wealth evolves according to the following stochastic process
	\begin{equation}\label{Xtprocess}
		dX_t=\left(rX_t+\pi_t(\mu-r)-c_t + w b_t\right) dt+\sigma\pi_tdB_t.
	\end{equation}
	Once the agent annuitizes, he or she receives a fixed periodic payment until death, $t=T_D$. The periodic payment of an annuity purchasable by wealth $X$ is $kX$, for some constant $k > r$. Thus we define the objective function with two concave utilities $U_1$ and $U_2$ which will be discussed in next section
	\begin{equation}\label{obj}
		J(x\left|c, \tau\right)= \mathbb{E}\left[\int_0^{\tau\wedge T_D}e^{-r t}U_1(c_t,b_t)dt+\mathbf{1}_{\{\tau < T_D\}}\int_{\tau\wedge T_D}^{T_D}e^{-r t}U_2\left(kX_\tau^{x,c,b,\pi}\right)dt\right],
	\end{equation}	
	where
	\begin{itemize}
		\item[(i)] $X_t$ is the retiree's total wealth at time $t$ with $X_0=x$;
		\item[(ii)] $c_t$ ~is the consumption rate, $b_t$ is the labor rate, and $w_t$ is the wage rate;
		\item[(iii)] $\pi_t$ is the amount invested in the risky asset with price $S_t$;
		\item[(iv)]  $\tau$ ~~is the annuitization time; and
		\item[(v)] $T_D$ is the retiree's death time.
	\end{itemize}
	
	We assume the retiree's force of mortality, denoted as $\delta$, is constant and is independent of the Brownian motion. According to the argument in \citet{gerrard2012choosing}, we introduce the death rate $\delta$ and rewrite the objective function (\ref{obj}) as
	\begin{equation}\label{OBJ}
		J(x\left|c, \tau\right) = \mathbb{E}\Bigg[\int_0^{\tau}e^{-\rho t}U_1(c_t,b_t)dt + \frac{e^{-\rho \tau}}{\rho}U_2(kX_\tau^{x,c,b,\pi})\Bigg], \text{ for }\rho = r + \delta.
	\end{equation}
	The optimization problem is to determine the four control variables, $c_t$, $b_t$, $\pi_t$ and $\tau$ in such a way as to maximize the discounted expected utility in (\ref{OBJ}). 
	
	\begin{defn}(Admissible control): \label{Admi}
		The admissible control set $\mathbf{U}$ of $\mathcal{F}$-progressively measurable processes $(c_t, b_t, \pi_t)$, and stopping times set $\mathbf{S}$ are such that
		\begin{itemize}
			\item[1)] $J(x|\tau)\le\infty$ for all $(c_t, b_t, \pi_t)\in\mathbf{U}$ and $\tau\in\mathbf{S}$;
			\item[2)] the control terms $\left(c_t,b_t\right)$ satisfy the constraints $0\leq c_t$ and $0 \leq b_t \leq \bar b$;
		\end{itemize}
	\end{defn}

	Definition \ref{Admi} ensures the bounded quadratic variation of the wealth process $X_t$ and the well-posedness of the control problem. Next, we derive the shadow constraint using the martingale approach. The value function is defined as the optimal objective function
	\begin{equation}\label{original}
		V(x)=\sup_{\substack{(c_t,b_t,\pi_t,\tau)\\\in \mathbf{U}\times\mathbf{S}}} \mathbb{E}\left[\int_0^{\tau}e^{-\rho t}U_1(c_t,b_t)dt + \frac{e^{-\rho \tau}}{\rho}U_2(kX_\tau^{x,c,b,\pi})\right].
	\end{equation}
	As is well known from the theory of stochastic control and optimal stopping (see, \citet{oksendal2003stochastic} and \citet{pham2009continuous}), the unconstrained value function can be characterized by an HJB equation. Therefore, we now proceed to derestrict the labor constraint from the original problem (\ref{original}) and obtain the dual problem by the martingale method. Define the relative risk process
	\begin{equation*}
		\theta = \frac{\mu - r}{\sigma},
	\end{equation*}
	the exponential martingale
	\begin{equation*}
		Z_t=e^{-\frac{1}{2}\theta^2t -\theta B_t},
	\end{equation*}
	and the state-price-density (shadow process)
	\begin{equation*}
		H_t=e^{-rt}Z_t=e^{-\left(r + \frac{1}{2}\theta^2\right)t -\theta B_t}.
	\end{equation*}
	Applying It\^{o}'s formula to the product of the processes $H_t$ and $X_t + \frac{w}{r}$, we have
	\begin{equation}\label{unconstrain_eqn}
		H_t \left(X_t + \frac{w}{r}\right) + \int_0^t {H_s\left(c_s + w(1-b_s)\right)ds } = x + \frac{w}{r} + \int_0^t{H_s\left(\sigma\pi_s - \theta \left( X_s + \frac{w}{r}\right)\right)dB_s}.
	\end{equation}
	By adding the term $\frac{w}{r}$, the process
	\begin{equation*}
		H_t \left(X_t + \frac{w}{r}\right) + \int_0^t {H_s\left(c_s + w(1-b_s)\right)ds },
	\end{equation*}
	is a continuous, positive local martingale, hence a supermartingale, under measure $\mathbb{P}$ for any feasible control set $(c_t,b_t,\pi_t)$. Therefore, by Fatou's Lemma the stochastic integral of equation (\ref{unconstrain_eqn}) is a $\mathbb{P}$-supermartingale. For simplicity, we denote $l_t = 1 - b_t$ and see it as the leisure rate with $\bar l \leq l_t \leq 1$ where $\bar l = 1 - \bar b \leq 1$ is the minimum leisure rate. By the optional sampling theorem, we have the following inequality constraint
	\begin{equation}\label{budget}
		\mathbb{E}\Bigg[\int_0^\tau {H_t\left(c_t + wl_t\right) dt} + H_\tau \left(X_\tau + \frac{w}{r}\right) \Bigg] \leq x + \frac{w}{r}, ~~~~~\forall \tau\in\mathbf{S}.
	\end{equation}
	We call the inequality (\ref{budget}) the budget constraint under which the solvency property $X_t^{c,l,\pi} \geq 0$ over the time interval $[0, \tau]$ is satisfied. In the special case when $\bar l = 1$, the agent receives no extra labor income after retirement and the model becomes a classical consumption-investment problem. If the labor rate $b_t$ is greater than $0$, then the agent has extra labor income $w_tb_t = w_t\left(1-l_t\right)$ with wage rate $w_t$ and he receives a penalty when $l_t < 1$. In the following sections, we will show that equality can hold for the budget constraint and a martingale approach can relax the control terms from the objective function. Eventually, we can transform the control and stopping problem to a simple stopping problem.
	
	\section{The optimization problem and dual approach}~\label{sec:3}

	In the classical \citet{merton1971optimum} problem, the optimal control term is obtained by solving the HJB equation which is reduced to a PDE that involves the first order and second order derivatives of the value function. If we follow this approach then the resulting PDE is difficult to solve due to labour constraints. Another approach to stochastic control/optimal stopping problems is the martingale method based on duality. The duality approach was employed by \citet{jin2006disutility}, however, an analytic solution is still difficult to find and the solution obtained this way only satisfies the necessary conditions for optimality. In this section, we first specify the utility functions, and then transform our stochastic control/optimal stopping problem into a dual optimal stopping problem. The dual problem is solved via an HJB variational inequality and this in turn is shown to yield the optimal solution.

	\subsection{Utility and conjugate form}
	Substituting the leisure rate $l_t = 1 - b_t$, we could consider measuring the utility of the leisure term instead of the dis-utility of the labor term and thus we could find a proper utility to measure them together. \citet{gerrard2012choosing} uses a quadratic function to measure the deviation of the consumption from a targeted value which is tractable for analysis but not appropriate for constructing the problem since higher consumption will also be penalized. We will consider a concave, monotone increasing, and isoelastic function as the utility function. One choice is the Cobb-Douglas utility function which is introduced in \cite{cobb1928theory} and has a decreasing marginal rate of substitution and a constant utility elasticity. The Cobb-Douglas utility function is often used in economics for analyzing of the relationship between capital and labor inputs. By the similarity between the two terms, we replace the capital and labor terms by consumption and leisure terms in the Cobb-Douglas utility and add two elasticity parameters $(\alpha,\beta)$:
	\begin{equation*}
		U_1(c,b) = U_1(c,l) = \frac{l^{\beta(1-p_1)} c^{\alpha(1-p_1)}}{1-p_1}.
	\end{equation*}
	After annuitization the leisure rate is $l=1$, so $U_1(\cdot)$ becomes a power utility
	\begin{equation*}
		U_2(x) = \frac{x^{1-p_2}}{1-p_2},
	\end{equation*}
	where $p_1, p_2 > 1$ are the coefficients of relative risk aversion and $\alpha,~\beta >0$ are elasticities of consumption and leisure.
	
	We can easily check that $U_1$ and $U_2$ satisfy elasticity requirement for utility functions. Since $l_t$ represents leisure, the fact $\frac{\partial U_1}{\partial \beta}=\left(l^\beta c^\alpha\right)^{-p_1} l^{\beta} \ln{l} <0$ shows that as the value of $\beta$ increases, the retiree receives less utility from the same amount of leisure which means he is willingly to work less. As the value of $\alpha$ increases, the retiree prefers to consume more. Therefore, we say $\beta$ is the retiree's labor preference and $\alpha$ is his consumption preference.  Next, we derive the conjugate utility using arguments similar to  \citet{barucci2012optimal}. 
	
	Applying Legendre-Fenchel transform for concave function, we define the conjugate function $\bar{U}_1$ and $\bar{U}_2$ for $U_1$ and $U_2$ as
	\begin{equation}\label{U1bar}
		\bar{U}_1(y) = \sup_{\substack{0\leq c\\ \bar l\leq l\leq 1}}\left[U_1(c,l)-(c + wl)y\right],
	\end{equation}
	and
	\begin{equation}\label{U2bar}	
		\bar{U}_2(y) = \sup_{x\geq0}\left[U_2(kx)- xy\right].
	\end{equation}	
	To optimize $\left[U_1(c,l) - (c+wl)y\right]$ over positive $c$ and $l$, first order conditions will give the optimizer since the function is concave in both $c$ and $l$. For optimization of $\left[U_1(c,l) - (c+wl)y\right]$ over the constraint $l\in[\bar l, 1]$, we consider the following two cases
	\begin{remark}\label{defn_optiaml}
		Optimization rules and conditions for the utility function in two cases:
		\begin{itemize}
			\item[(i)] the optimizer of equation (\ref{U1bar}) is on the interior of the constraint set, or,
			\item[(ii)] the optimizer of equation (\ref{U1bar}) is on the boundary of the constraint set.
		\end{itemize}
	\end{remark}
	If case (i) in Remark \ref{defn_optiaml} holds, which means that $l\in[\bar l,1]$ and $y\in[\tilde{y},\bar{y}]$ in terms of $y$. We can derive the unconstrained optimizer of $c$ and $l$ from first order condition
	\begin{equation}\label{U1barFirstCond}
		\frac{1}{w}\frac{\partial \bar{U}_1}{\partial l} = \frac{\partial \bar{U}_1}{\partial c}=y,
	\end{equation}
	from equation (\ref{U1barFirstCond}) we get the unconstrained relationship between $c$ and $l$ when $l$ is unconstrained
	\begin{equation}\label{c_l_relation}
		c=\frac{\alpha w}{\beta}l,~\text{for }l\in[\bar l, 1].
	\end{equation}
	Replacing $c$ in the first order equation (\ref{U1barFirstCond}) using the unconstrained relationship (\ref{c_l_relation}), we obtain the following optimal condition for unconstrained $l$
	\begin{equation*}
		y = \alpha \left(\frac{\alpha w}{\beta}\right)^{\alpha\left(1-p_1\right)-1}{l}^{\left(\alpha+\beta\right)\left(1-p_1\right)-1},~\text{for }l\in[\bar l, 1].
	\end{equation*}
	Equivalently, we define the unconstrained region for $y$ as $[\tilde{y}, \bar{y}]$
	\begin{equation*}
		\tilde{y} = \alpha \left(\frac{\alpha w}{\beta}\right)^{\alpha\left(1-p_1\right)-1}\leq y \leq \alpha \left(\frac{\alpha w}{\beta}\right)^{\alpha\left(1-p_1\right)-1}{\bar l}^{\left(\alpha+\beta\right)\left(1-p_1\right)-1} = \bar{y},
	\end{equation*}
	and the unconstrained region for $c$ as $[\bar{c}, \tilde{c}]$
	\begin{equation*}
		\bar{c} = \frac{\alpha w}{\beta}\bar l\leq c \leq \frac{\alpha w}{\beta} = \tilde{c}.
	\end{equation*}
	If case (i) in Remark \ref{defn_optiaml} fails, then $y\notin[\tilde y, \bar{y} ]$ and case (ii) in Remark \ref{defn_optiaml} gives $y<\tilde{y}$ and $c>\tilde{c}$ or $y>\bar{y}$ and $c<\bar{c}$. Denote by $l^*(c) =\frac{\beta c}{\alpha w} $, we rewrite equation (\ref{U1bar}) as 
	\begin{align}\label{reduced_U1bar}
		\bar{U}_1(y) 
		= &\left\{U_1(c,\bar l)-(c + w\bar l)y\right\}\mathbf{1}_{y>\bar{y}} \nonumber\\ 
		&+\left\{U_1(c,l^*(c))-(c + w l^*(c))y\right\}\mathbf{1}_{\tilde{y}\leq y\leq \bar{y}}+
		\left\{U_1(c,1)-(c + w)y\right\}\mathbf{1}_{y<\tilde{y}}.
	\end{align}
	If we take the partial derivative with respect to $c$ in equation (\ref{reduced_U1bar}), we get the following system
	\begin{equation}\label{C_system}
		\left\{
		\begin{alignedat}{2}
			&\frac{\partial }{\partial c} U_1(c,\bar l)-y  = 0,~~&&y > \bar{y}\\
			&\frac{\partial }{\partial c} U_1(c,l^*(c))-(1 + \frac{\alpha}{\beta})y = 0,~~&&\tilde{y}\leq y \leq \bar{y}\\
			&\frac{\partial }{\partial c} U_1(c,1)-y = 0,~~&&y<\tilde{y},
		\end{alignedat}
		\right.
	\end{equation}
	which solves the optimal consumption on three regions: the full labor region $(0,\tilde{y})$ for $c> \tilde{c}$, the full leisure region $(\bar{y},\infty)$ for $0<c<\bar{c}$, and the flexible labor region $[\tilde{y},\bar{y}]$ for $c\in[\bar{c},\tilde{c}]$.	
	
	Solving each equation for $c$ in equation (\ref{C_system}), we get the optimal value of $c$
	\begin{equation}\label{optimal_c}
		\left\{
		\begin{alignedat}{2}
			&c=\left(y/\alpha\right)^{p_1'-1}\bar l^{-\frac{\beta}{\alpha}p_1'},~~ &&  y > \bar{y}\\
			&c=\left(y/\alpha\right)^{p-1}\left(\alpha w/\beta\right)^{\frac{\beta}{\left(\alpha+\beta\right)}p},~~ &&  \tilde{y}\leq y \leq \bar{y}\\
			&c=\left(\frac{y}{\alpha}\right)^{p_1'-1},~~ && y<\tilde{y},
		\end{alignedat}
		\right.
	\end{equation}
	where
	\begin{align}
		p =& \frac{\left(\alpha+\beta\right)\left(1-p_1\right)}{\left(\alpha+\beta\right)\left(1-p_1\right)-1} \in(0,1),\label{p_val}\\
		p_1' =& \frac{\alpha\left(1-p_1\right)}{\alpha\left(1-p_1\right)-1}\in(0,1).\label{p1_val}
	\end{align}
	Therefore, equation (\ref{reduced_U1bar}) can be simplified as
	\begin{equation}\label{U1bar_eqn}
		\bar{U}_1(y) = \Bigg[\tilde{A} y^{p_1'} - wy\Bigg]\mathbf{1}_{\{0<y\leq \tilde{y}\}} + A y^p \mathbf{1}_{\{\tilde{y} < y< \bar{y}\}} + \Bigg[\bar{A} y^{p_1'} - w\bar l y\Bigg]\mathbf{1}_{\{\bar{y}\leq y \}},
	\end{equation}
	where $A = -\frac{\alpha+\beta}{p}\alpha^{-\frac{\alpha}{\alpha + \beta}}\beta^{-\frac{\beta}{\alpha + \beta}}w^{\frac{\beta}{\alpha+\beta}p} <0$, $\tilde{A} = -\frac{\alpha^{1-p_1'}}{p_1'} <0$ and $\bar{A} = -\frac{\alpha^{1-p_1'}}{p_1'}\bar l^{-\frac{\beta}{\alpha}p_1'} <0$.

	The optimal value of $c$ and $l$ are given by $c^* = I_c(y)$ and $l^* = I_l(y)$
	\begin{align}
		I_c(y)=&\left(\frac{y}{\alpha}\right)^{p_1'-1}\mathbf{1}_{\{0< y\leq \tilde{y}\}} +  \left(\frac{y}{\alpha}\right)^{p-1}\left(\frac{\alpha w}{\beta}\right)^{\frac{\beta}{\left(\alpha+\beta\right)}p} \mathbf{1}_{\{\tilde{y} < y< \bar{y}\}}\notag\\
		& + \left(\frac{y}{\alpha}\right)^{p_1'-1}\bar l^{-\frac{\beta}{\alpha}p_1'}\mathbf{1}_{\{\bar{y}\leq y\}},\label{I_c}\\
		I_l(y)=& \mathbf{1}_{\{0< y\leq \tilde{y}\}} +  \left(\frac{y}{\alpha}\right)^{p-1}\left(\frac{\alpha w}{\beta}\right)^{\frac{\beta}{\left(\alpha+\beta\right)}p-1}\mathbf{1}_{\{\tilde{y} < y< \bar{y}\}} + L \mathbf{1}_{\{\bar{y}\leq y\}},\label{I_l}
	\end{align}
	where $I_c(\cdot)$ is the inverse of the marginal utility of consumption $\frac{\partial U_1}{\partial c}$ and $I_l(\cdot)$ is the inverse of the marginal utility of labor $\frac{\partial U_1}{\partial l}$. We can easily show that $I_c$ and $I_l$ are decreasing functions; the function $I_c$ maps $(0,\infty)$ onto itself and satisfies $I_c(0^+)=+\infty$ and $I_c(+\infty) = 0$; the function $I_l$ satisfies $I_l(0^+) = 1$ and $I_l(+\infty) = \bar l$.
	
	The conjugate function $\bar{U}_2(y)$ is given by
	\begin{equation*}
		\bar{U}_2(y)=U_2(kx^*)- x^*y=-\frac{1}{p_2'}\left(\frac{y}{k}\right)^{p_2'},
	\end{equation*}
	where 
	\begin{equation}\label{p2_val}
		p_2' = \frac{p_2 - 1}{p_2}\in(0,1).
	\end{equation}
	The optimal values $x^*$ and he marginal utility $I(y)$ are given by
	\begin{equation}\label{I_y}
		x^* = I(y) =  \frac{1}{k}\left(\frac{y}{k}\right)^{p'_2 - 1}.
	\end{equation}
	From the definition of the conjugate functions (\ref{U1bar}) and (\ref{U2bar}) and inverse marginal functions $I_c(\cdot)$, $I_l(\cdot)$ and $I(\cdot)$, we have
	\begin{align*}
		\bar{U}'_1(y) &=-\left(I_c(y) + wI_l(y)\right),\\
		\bar{U}'_2(y) &= -I(y),
	\end{align*}
	and
	\begin{equation} \label{U2bar_eqn}
		U_2(kx) = \min_{y>0}\left[\bar{U}_2(y) + xy\right] = \bar{U}_2\left(I'(x)\right) + xI'(x).
	\end{equation}
	It can be shown that $\bar{U}_1(\cdot)$ and $\bar{U}_2(\cdot)$ are strictly decreasing and convex.
	
	Following the conjugate transform, we obtain the optimal values for $c^*$, $l^*$ and $x^*$ depending on some $y$. In next section, we will specify the relationship between $x$ and $y$ such that the controls $c^*$, $l^*$ and $x^*$ are optimal for the original problem.
	
	\subsection{Martingale and convex duality methods}
	Inspired by the martingale methodology used in \citet{choi2008optimal}, we consider a state process that could carry out the optimization and obtain a dual problem without the restrictions from the control processes. We consider the dual of the wealth process $X_t$ by means of a series of super-martingales satisfying the budget constant (\ref{budget}) at the same time. We define the dual of $X_t$ as the shadow process $Y_t^y$
	\begin{equation}\label{Y_process}
		Y_t^y=ye^{\rho t} H_t.
	\end{equation}
	with the stochastic dynamics
	\begin{equation*}
		dY_t = (\rho-r)Y_t dt - \theta Y_t dB_t.
	\end{equation*}
	It is easy to check that shadow process $Y_t$ is inversely proportional to the wealth process $X_t$, that is, when $X_t$ increases $Y_t$ decreases.
	 
	For a Lagrange multiplier $y > 0$, we give the dual objective function as the Lagrange function of (\ref{OBJ}) and its budget constraint (\ref{budget})
	\begin{equation}\label{dual_eqn2}
		\bar{J}(x,y|\tau)=J(x|\tau) + y\left(\left(x+\frac{w}{r}\right)-\mathbb{E}\left[\int_0^\tau { H_t(c_t  + wl_t)dt} + H_\tau \left(X_\tau + \frac{w}{r}\right)\right]\right).
	\end{equation}
	By the inequality of budget constraint (\ref{budget}), it is obvious that
	\begin{equation}\label{OBJ_inequality}
		\bar{J}(x,y|\tau) \geq J(x|\tau).
	\end{equation}
	
	We replace the conjugate utilities in equation (\ref{dual_eqn2}) by their explicit forms defined in (\ref{U1bar_eqn}) and (\ref{U2bar_eqn}), then we obtain the following inequality
	\begin{align}\label{dual_OBJ}
		\sup_{c_t,l_t}\bar{J}(x,y|\tau)=&\sup_{c_t,l_t}\mathbb{E}\left[\int_0^{\tau}e^{-\rho t}\left(U_1\left(c_t,l_t\right)-yH_t\left(c_t  + wl_t\right)\right)dt \right.\nonumber\\
		& ~~~~~~~~~~~~~~~+ \left.\left(\frac{e^{-\rho \tau}}{\rho}U_2(kX_\tau) - yH_\tau \left(X_\tau + \frac{w}{r}\right)\right)\right] + y\left(x+\frac{w}{r}\right)\nonumber\\
		\leq&\mathbb{E}\left[\int_0^{\tau}e^{-\rho t}\sup_{c_t,l_t}\left(U_1\left(c_t,l_t\right)-Y_t^y\left(c_t  + wl_t\right)\right)dt \right.\nonumber\\
		& ~~~~~~~~~~~~~~~+ \left.\sup_{c_t,l_t}e^{-\rho \tau}\left(\frac{1}{\rho}U_2(kX_\tau) - Y_\tau^y \left(X_\tau + \frac{w}{r}\right)\right)\right] + y\left(x+\frac{w}{r}\right)\nonumber\\
		=&\mathbb{E}\Bigg[\int_0^{\tau} e^{-\rho t}\bar{U}_1(Y_t^y)dt + e^{-\rho \tau}\left(\frac{1}{\rho}\bar{U}_2(\rho Y_\tau^y) -\frac{w}{r}Y_\tau^y \right)\Bigg] + y\left(x+\frac{w}{r}\right).
	\end{align}
	Actually the inequality in (\ref{dual_OBJ}) can be equality under the optimal control functionals given in (\ref{optimal_c}), and we could define the dual objective function as
	\begin{equation*}
		\bar V(x,y|\tau) = \sup_{c_t,l_t}\bar{J}(x,y|\tau)=\mathbb{E}\Bigg[\int_0^{\tau} e^{-\rho t}\bar{U}_1(Y_t^y)dt + e^{-\rho \tau}\left(\frac{1}{\rho}\bar{U}_2(\rho Y_\tau^y) -\frac{w}{r}Y_\tau^y \right)\Bigg] + y\left(x+\frac{w}{r}\right).
	\end{equation*}
	Hence, the dual problem becomes an optimal stopping problem
	\begin{equation}\label{dual}
		\bar{V}(x,y)=\sup_{\tau}\bar{V}(x,y|\tau)
		=\sup_{\tau}\mathbb{E}\Bigg[\int_0^{\tau} e^{-\rho t}\bar{U}_1(Y_t^y)dt + e^{-\rho \tau}\left(\frac{1}{\rho}\bar{U}_2(\rho Y_\tau^y) -\frac{w}{r}Y_\tau^y \right)\Bigg] + y\left(x+\frac{w}{r}\right).
	\end{equation}
	As the conjugate utility $\bar U_1$ and $\bar U_2$ are strictly decreasing functions, we would expect a boundary value $y^*>0$ and a stopping time $\tau^*_y=\inf\left\{t:Y_t^y \leq y^*\right\}$ to solve the dual problem (\ref{dual}) if the supremum is obtainable.

	Before reaching the solution to the dual problem, we consider the relationship between the original problem and the dual problem. By equation (\ref{OBJ_inequality}) we obtain the following inequality between the original and dual objective functions
	\begin{equation}\label{Dual_OBJ}
		V(x|\tau) = \sup_{(c_t,l_t,\pi_t)} J(x|\tau)\leq \sup_{(c_t,l_t,\pi_t)} \bar{J}(x,y|\tau)=\bar{V}(x,y|\tau) \leq \sup_{\tau}\bar{V}(x,y|\tau) = \bar{V}(x,y).
	\end{equation}
	The second inequality in (\ref{Dual_OBJ}) always holds for all $x>-\frac{w}{r}$ and $y>0$, so we obtain the following inequalities
	\begin{equation}\label{ineq0}
		V(x|\tau) \leq \inf_{y>0}\bar{V}(x,y|\tau)\leq \inf_{y>0}\bar{V}(x,y),
	\end{equation}
	and
	\begin{equation}\label{ineq1}
		V(x) =\sup_{\tau}V(x|\tau)\leq \sup_{\tau}\inf_{y>0} \bar{V}(x,y|\tau) = \inf_{y>0} \bar{V}(x,y).
	\end{equation}
	By observing the two inequalities in (\ref{ineq1}), we can reach following two conclusions:
	\begin{itemize}
		\item[1.] If the first inequality in (\ref{Dual_OBJ}) is equality for some $\hat y$, then the control terms $c^*_t = I_c\left(Y_t^{\hat y}\right)$ and $l^*_t = I_l\left(Y_t^{\hat y}\right)$ obtained from the dual value function (\ref{dual}) is the optimal controls to the original problem
		\item[2.] If the second inequality in (\ref{Dual_OBJ}) is equality for some $\tau^*_y$, then we can obtain the optimal stopping time for the original problem as well.
	\end{itemize}
	
	Following above ideas, we find a precise connection between the original problem and the dual problem. In fact, the first inequality in (\ref{Dual_OBJ}) can become equality as long as there exists an $\pi_t$ that satisfies the budget constraint
	\begin{equation}\label{budgethold}
		x + \frac{w}{r} = \mathbb{E}\left[\int_0^\tau { H_t\left(I_c(Y_t^y) + w I_l(Y_t^y)\right)dt} + H_\tau \left(I(\rho Y_\tau^y) + \frac{w}{r}\right)\right].
	\end{equation}
	We summarize the above in the following lemma.
	\begin{lemma}[Obtainable budget constraint]\label{equality_lemma}
		For any $\tau$, any $\mathcal{F}_\tau$-measurable $B$ with $\mathbb{P}\left[B>0\right]=1$, any progressively measurable process $c_t\geq 0$ and $l_t \in[0,1]$ that satisfy, for all $t\leq \tau$
		\begin{equation*}
			\mathbb{E}\left[\int_0^\tau {H_t\left(c_t + wl_t\right) dt} + H_\tau \left(B +\frac{w}{r}\right) \right] = x + \frac{w}{r},
		\end{equation*}
		there exists a portfolio process $\pi_t$ such that, a.e.	
		\begin{equation*}
			X_t^{\left(c,l,\pi\right)} > -\frac{w}{r},~
			\text{ for all $0\leq t <\tau$, and }
			~X_\tau^{\left(c,l,\pi\right)} = B.
		\end{equation*}
	\end{lemma}
	\noindent Proof: see Appendix \ref{Pf_lemma_equality_lemma}.
	\bigskip

	\noindent Before moving on with the optimal dual problem, we need to check that all $x$ and $y$ that satisfy (\ref{budgethold}) cover $(-\frac{w}{r},\infty)$ and $(0,\infty)$. To see this, we define the following map on $(0,\infty)$,
	\begin{equation}\label{XmapY}
		\mathbb{X}_\tau (y) = \mathbb{E}\left[\int_0^\tau { H_t\left(I_c(Y_t^y) + w I_l(Y_t^y)\right)dt} + H_\tau \left(I(\rho Y_\tau^y) + \frac{w}{r}\right)\right] - \frac{w}{r},~\text{for $\tau\in\mathbf{S}$},
	\end{equation}
	Recall from equations (\ref{I_c}), (\ref{I_l}) and (\ref{I_y}) that $I_c(\cdot)$, $I_l(\cdot)$ and $I(\cdot)$ are continuous, strictly decreasing functions satisfying $I_c(0^+) = I(0^+)=+\infty$ and $I_c(+\infty) = I(+\infty)= 0$. We can verify that $\mathbb{X}(y)$ is a continuous, strictly decreasing function with $\mathbb{X}(0^+) = \infty$ and $\mathbb{X}(\infty)>0$. Therefore, $\mathbb{X}_\tau(y)$ is a map from $(0,\infty)$ onto $(0,\infty)$ for all $\tau\in\mathbf{S}$. Thus $\mathbb{X}_\tau$ has an one-to-one inverse function $\mathbb{Y}_\tau(\cdot)$ from $(0,\infty)$ onto $(0,\infty)$ with $\mathbb{Y}_\tau(0) = \infty$ and $\mathbb{Y}_\tau(\infty) >0$. For any $\tau\in\mathbf{S}$, and all $y\in(0,\infty)$
	\begin{equation}\label{set_1}
		\bigcup\limits_{\tau\in\mathbf{S},y\in(0,\infty)}\mathbb{X}_\tau(y),
	\end{equation}
	covers $(0,\infty)$. For any $\tau\in\mathbf{S}$, and all $x\in(0,\infty)$
	\begin{equation}\label{set_2}
		\bigcup\limits_{\tau\in\mathbf{S},x\in(0,\infty)}\mathbb{Y}_\tau(x),
	\end{equation}
	covers $(0,\infty)$. Therefore, the two sets (\ref{set_1}) and (\ref{set_2}) have no gap between initial $x$ in original problem and initial $y$ in dual problem. Therefore, we conclude that the inequality in (\ref{ineq1}) can become an equality. We present the following result given the smoothness of the dual function.
	\begin{lemma}[Optimal dual function]\label{Vbar_prime}
		If $\bar{V}(x,y)$ of (\ref{dual}) is differentiable at $y > 0$, then
		\begin{equation}
			\frac{\partial \bar{V}}{\partial y}(x,y) = 0.
		\end{equation}
	\end{lemma}
	\noindent Proof: see Appendix \ref{Pf_lemma_Vbar_prime}.
	\bigskip

	\noindent By Lemma \ref{equality_lemma} and the inequality in (\ref{Dual_OBJ}), we can obtain the optimal solution from the following theorem.
	\begin{theorem}[Attainable dual problem]\label{equality_thm}
		If the value function $V(x)$ of problem (\ref{original}) is attainable, then the dual function $\bar V(x,y)$ of the problem (\ref{dual}) is also attainable, such that
		\begin{equation}\label{link_function}
			V(x) = \inf_{y >0}\bar{V}(x,y),
		\end{equation}
		and vice versa.
	\end{theorem}
	\noindent Proof: see Appendix \ref{Pf_theorem_euqality_thm}.
	\bigskip
	
	It remains to solve the dual problem (\ref{dual}) and then we can find the solution to the original problem (\ref{original}) by Lemma \ref{Vbar_prime} and Theorem \ref{equality_thm}. Following the dynamic programming principle, we consider the stopping problem on a small interval $[t,\tau]$
	\begin{equation}\label{OSP}
		\psi\left(t,y\right)=\sup_{\tau > t}\mathbb{E}\Bigg[\int_t^{\tau} e^{-\rho s}\bar{U}_1(Y_s)ds + e^{-\rho \tau}\left(\frac{1}{\rho}\bar{U}_2(\rho Y_\tau) -\frac{w}{r}Y_\tau \right)\Big|Y_t=y\Bigg].
	\end{equation}
	Define the differential operator $\mathcal{L} = (\rho - r)y\frac{\partial}{\partial y} + \frac{1}{2}\theta^2 y^2 \frac{\partial^2}{\partial y^2}$, then we obtain the following HJB equation (see \citet{fleming2012deterministic})
	\begin{equation}\label{HJB}
		\max\left\{e^{-\rho t}\left(\frac{1}{\rho}\bar{U}_2(\rho y) -\frac{w}{r}y \right) - \psi(t,y),~~ \frac{\partial \psi}{\partial t} + \mathcal{L}\psi + e^{-\rho t}\bar{U}_1(y)\right\} = 0.
	\end{equation}
	Introducing the ansatz $\psi(t,y) = e^{-\rho t}\phi(y)$ to separate the time term from the objective function, we rewrite the HJB equation (\ref{HJB}) as 
	\begin{equation}\label{HJB*}
		\max\left\{\frac{1}{\rho}\bar{U}_2(\rho y) -\frac{w}{r}y - \phi(y),~~ -\rho \phi(y) + \mathcal{L}\phi(y) + \bar{U}_1(y)\right\} = 0.
	\end{equation}
	By analogy to \citet[Sec. 2.7]{karatzas1998}, the solution to the HJB equation (\ref{HJB*}) satisfies the following variational inequalities. We determine the boundary ${y^*}$ and a continuous function $\phi \in C^1\left(\mathbb{R}^+\right)\cap C^2\left( \mathbb{R}^+\setminus \{{y^*}\}\right)$ satisfying:
	\begin{defn}(Variational inequality)
		Find a free boundary ${y^*}$ and a non-increasing convex function $\phi \in C^1\left(\mathbb{R}^+\right)\cap C^2\left( \mathbb{R}^+\setminus \{{y^*}\}\right)$ satisfying
		\begin{equation}\label{Variational}
			\left\{
			\begin{alignedat}{2}
				&-\rho \phi + \mathcal{L}\phi + \bar{U}_1(y)  = 0, ~~&&{y^*} < y\\
				&-\rho \phi + \mathcal{L}\phi + \bar{U}_1(y) \leq 0, ~~&&0 < y \leq {y^*} \\
				&\phi(y) \geq \frac{1}{\rho}\bar{U}_2(\rho y) - \frac{w}{r}y, ~~&&{y^*} < y\\
				&\phi(y) = \frac{1}{\rho}\bar{U}_2(\rho y) - \frac{w}{r}y, ~~&&0 < y \leq {y^*}.
			\end{alignedat}
			\right.
		\end{equation}
	\end{defn}
	\noindent If $\phi(y)$ solves the variational inequalities (\ref{Variational}) then according to the definition of the dual problem (\ref{dual}) the solution to the dual problem is given by 
	\begin{equation*}
		\bar{V}(x,y) = \psi(0, y) + y\left(x+\frac{w}{r}\right) = \phi(y) + y\left(x+\frac{w}{r}\right),
	\end{equation*}
	thus we have the following verification theorem as a result of (\ref{dual}) and (\ref{Variational}).
	\begin{theorem}[Verification theorem]\label{Verification}
		Suppose $\phi(y)$ is a solution to (\ref{Variational}) and there exists a ${y^*}$ such that $\phi'(y)$ is absolutely continuous at ${y^*}$. Then $\bar{V}(x,y) = \phi(y)+ y\left(x+\frac{w}{r}\right)$ and $\tau^*_y = \inf \{t\geq 0: Y_t^y \leq {y^*}\}$ are the unique solution to the dual problem (\ref{dual}).
	\end{theorem}
	\noindent Proof: see Appendix \ref{Pf_theorem_Verification}.
	\bigskip
	
	\noindent The following two lemmas provide the unique solution to the variational inequalities.
	\begin{lemma}[Existence and uniqueness]\label{existence}
		If $p'_1 < p'_2$ then there exists an unique ${y^*} >0$ such that
		\begin{equation}\label{exist_eqn}
			\int_{+\infty}^{{y^*}} {\frac{\bar{U}_1(z) + \frac{n(p'_2)}{\rho}\bar{U}_2(\rho z) + wz}{z^{n_1 +1}}}dz = 0,
		\end{equation}
		where $n(x) = \frac{1}{2}\theta^2 x^2 + (\rho - r - \frac{1}{2}\theta^2)x -\rho$.
	\end{lemma}
	\noindent Proof: see Appendix \ref{Pf_lemma_existence}.
	\bigskip
	
	\begin{lemma}[Optimal solution]\label{uniqueness}
		If $p'_1 < p'_2$ then there exists an unique free boundary ${y^*} >0$ and an unique function $\phi$ that solves the free boundary ODE in (\ref{Variational}).
		\begin{equation*}
			\phi(y) = \left\{
			\begin{alignedat}{2}
				& C y^{n_2} + \frac{2y^{n_1}}{\theta^2(n_1 - n_2)}\int_{+\infty}^y {\frac{-\bar{U}_1(z)}{z^{n_1 +1}}}dz - \frac{2y^{n_2}}{\theta^2(n_1 - n_2)}\int_{{y^*}}^y {\frac{-\bar{U}_1(z)}{z^{n_2 +1}}}dz, &&~~~~{y^*} < y \\
				& \frac{1}{\rho}\bar{U}_2(\rho y) - \frac{w}{r}y, &&~~~~0 < y \leq {y^*},
			\end{alignedat}
			\right.
		\end{equation*}
		where
		\begin{equation*}
			C =\frac{{{y^*}}^{-n_2}}{n_1-n_2}\left((n_1-p'_2)\frac{1}{\rho}\bar{U}_2(\rho{y^*}) -(n_1-1) \frac{w}{r}{y^*}\right),
		\end{equation*}
		and $n_1$ and $n_2$ are the two roots of
		\begin{equation*}
			n(x)=\frac{1}{2}\theta^2 x^2 + (\rho - r - \frac{1}{2}\theta^2)x -\rho ,
		\end{equation*}
		with $n_1 > 1$, $n_2 \leq -\frac{2\left(\rho - r\right)}{\theta^2}$.
	\end{lemma}
	\noindent Proof: see Appendix \ref{Pf_lemma_uniqueness}.
	\bigskip
	
	Consider the case of no labor income $(\bar b=0)$. By Lemma \ref{uniqueness} we have
	\begin{equation*}
		\frac{2\bar{A} {{y^*}}^{p_1'}}{\theta^2(p_1'- n_1)}= \frac{(p'_2 - n_2){{y^*}}^{p_2'}}{\rho p'_2}\left(\frac{\rho}{k}\right)^{p_2'}.
	\end{equation*}
	The value $p_1'=\frac{\alpha\left(1-p_1\right)}{\alpha\left(1-p_1\right)-1}$ and $p_2' = \frac{p_2-1}{p_2}$ given in equations (\ref{p1_val}) and (\ref{p2_val}) are the same in the models with labor income or without labor income. The optimal solution to the model without labor exists if and only if $p'_1\neq p'_2$ and the unique solution to the dual value function is given by
	\begin{equation*}
		{y^*} = \left(-\frac{p_1'}{p_2'}\frac{\theta^2}{2\rho}(p_1'-n_1)(p_2'-n_2)\alpha^{p_1'-1}\left(\frac{\rho}{k}\right)^{p_2'}\right)^{\frac{1}{p_1'-p_2'}}.
	\end{equation*}
	According to the Lemma \ref{uniqueness}, we define the critical wealth level
	\begin{equation*}
		{x^*} =I(\rho {y^*}).
	\end{equation*}
	By Lemma \ref{Vbar_prime}, Theorem \ref{equality_thm}, Theorem \ref{Verification} and Lemma \ref{uniqueness}, we have the following theorem which gives the main results of the original value function.
	\begin{theorem}[Value function]\label{Optimal_thm}
		If $\alpha(1-p_1) > 1-p_2$, then there exists an unique free boundary ${x^*}$ that solves the optimal stopping problem (\ref{original}) and the value function $V(x)$ is given by
		\begin{equation*}
			V(x) = 
			\begin{cases}
				y\left(x + \frac{w}{r}\right) + C y^{n_2} + \frac{2y^{n_1}}{\theta^2(n_1 - n_2)}\int_{+\infty}^y {\frac{-\bar{U}_1(z)}{z^{n_1 +1}}}dz  
                                - \frac{2y^{n_2}}{\theta^2(n_1 - n_2)}\int_{{y^*}}^y {\frac{-\bar{U}_1(z)}{z^{n_2 +1}}}dz, & 0<x <{x^*},\\
				\frac{1}{\rho}U_2(x), &{x^*}\leq x,
			\end{cases}
		\end{equation*}
		where $y$ is a function of $x$ which is determined from the following equation
		\begin{align}\label{xofy}
			x = &- C n_2 {y}^{n_2-1} - \frac{2n_1{y}^{n_1-1}}{\theta^2(n_1 - n_2)}\int_{+\infty}^{y} {\frac{-\bar{U}_1(z)}{z^{n_1 +1}}}dz + \frac{2n_2{y}^{n_2-1}}{\theta^2(n_1 - n_2)}\int_{{y^*}}^{y} {\frac{-\bar{U}_1(z) }{z^{n_2 +1}}}dz - \frac{w}{r}.
		\end{align}
		The optimal stopping boundary ${x^*}$ is given by
		\begin{equation}\label{xstar}
			{x^*} = I(\rho{y^*}),
		\end{equation}
		where ${y^*}$ is determined by the following equation
		\begin{equation}\label{ystar}
			\frac{2{y^*}^{n_1}}{\theta^2}\int_{+\infty}^{{y^*}} {\frac{\bar{U}_1(z) + wz}{z^{n_1 +1}}}dz = -(p_2'-n_2)\frac{1}{\rho}\bar{U}_2(\rho {y^*}).
		\end{equation}
	\end{theorem}
	
	Equation (\ref{xofy}) in Theorem \ref{Optimal_thm} shows the map between wealth level $x$ in the original problem (\ref{original}) and the shadow price $y$ in dual problem (\ref{dual}). To obtain the optimal wealth process and optimal strategy, we let $Y_t$ be the stochastic process defined in equation (\ref{Y_process}) and $y$ is the solution from the Theorem \ref{Optimal_thm}. We substitute $Y_t$ for $y$ and $X_t$ for $x$ into equation (\ref{xofy}), then we get the optimal wealth process
	\begin{align}\label{YtoX}
		X_t = - C n_2 {(Y_t)}^{n_2-1} &- \frac{2n_1{(Y_t)}^{n_1-1}}{\theta^2(n_1 - n_2)}\int_{+
			\infty}^{Y_t} {\frac{-\bar{U}_1(z)}{z^{n_1 +1}}}dz %
		&+ \frac{2n_2{(Y_t)}^{n_2-1}}{\theta^2(n_1 - n_2)}\int_{{y^*}}^{Y_t} {\frac{-\bar{U}_1(z) }{z^{n_2 +1}}}dz - \frac{w}{r}.
	\end{align}
	The optimal stopping time for both the original problem (\ref{original}) and the dual problem (\ref{dual}) is given by
	\begin{equation*}
		\tau^* = \inf\left\{t: Y_t^y \leq {y^*}\right\}.
	\end{equation*}
	Express equation (\ref{YtoX}) in terms of $\phi(y)$
	to see that
	\begin{equation*}
		\pi_{t}=\frac{\theta}{\sigma} Y_{t} \phi^{\prime\prime}\left(Y_{t}\right).
	\end{equation*}

	Through the proofs of Lemma \ref{existence} and Lemma \ref{uniqueness}, we see that the condition $\alpha(1-p_1)>1-p_2$ is a necessary and sufficient condition ensuring there exists a unique solution $y^*$ to the stopping problem. If this condition is violated, we could not obtain a solution to equation (\ref{exist_eqn}) and the optimal stopping problem is ruined when $\alpha(1-p_1)\leq1-p_2$. In the ruined problem, the retiree would continue the optimal consumption, labor, and investment strategies without choosing final annuitization. However, if the problem is ruined or not, the same optimal strategies for  consumption, labor, and investment will apply to the stopping problem and the ruined problem.
	\begin{corollary}[Ruined solution]\label{Nonstopping_corolarry}
		If $\alpha(1-p_1) \leq 1-p_2$, then there exists no optimal stopping boundary for (\ref{original}) and the value function $V(x)$ for $x>0$ is given by
		\begin{equation*}
			V(x) = y\left(x + \frac{w}{r}\right) + \frac{2y^{n_1}}{\theta^2(n_1 - n_2)}\int_{+\infty}^y {\frac{-\bar{U}_1(z)}{z^{n_1 +1}}}dz- \frac{2y^{n_2}}{\theta^2(n_1 - n_2)}\int_0^y {\frac{-\bar{U}_1(z)}{z^{n_2 +1}}}dz,
		\end{equation*}
		where $y$ is determined by $x$ from equation (\ref{xofy}).
	\end{corollary}

	\begin{theorem}[Optimal controls]\label{strategy}
		The optimal strategies $(c^*_t, b^*_t, \pi^*_t, \tau^*)$ are given by
		\begin{align*}
			c_t^* =& \left\{
			\begin{alignedat}{2}
				&\left(\frac{Y_t}{\alpha}\right)^{p'_1-1},	~~~~~~~~~~~~&& \mbox{if } Y_t <\tilde{y}\\
				&\left(\frac{Y_t}{\alpha}\right)^{p-1}\left(\frac{\alpha w}{\beta}\right)^{\frac{\beta}{\left(\alpha+\beta\right)}p}, ~~~~~~~~~~~~&&\mbox{if } \tilde{y} \leq Y_t \leq \bar{y}\\
				&\left(\frac{Y_t}{\alpha}\right)^{p'_1-1}\left(1-\bar b\right)^{-\frac{\beta}{\alpha}p'_1}, ~~~~~~~~~~~~&&\mbox{if } \bar{y}<Y_t,
			\end{alignedat}
			\right.\\
			b_t^* =& \left\{
			\begin{alignedat}{2}
				&0,	~~&&\mbox{if } Y_t < \tilde{y}\\
				&1-\left(\frac{Y_t}{\alpha}\right)^{p-1}\left(\frac{\alpha w}{\beta}\right)^{\frac{\beta}{\left(\alpha+\beta\right)}p -1}, ~~&&\mbox{if } \tilde{y} \leq Y_t\leq \bar{y}\\
				&\bar b, ~~&&\mbox{if } \bar{y} < Y_t,
			\end{alignedat}
			\right.\\
			\pi_t^* =& \frac{2n_1(n_1-1)Y_t^{n_1-1}}{\theta^2(n_1 - n_2)}\int_{+\infty}^{Y_t} {\frac{-\bar{U}_1(z)}{z^{n_1 +1}}}dz - \frac{2n_2(n_2-1)Y_t^{n_2-1}}{\theta^2(n_1 - n_2)}\int_{{y^*}}^{Y_t} {\frac{-\bar{U}_1(z) }{z^{n_2 +1}}}dz\\
			&+ \frac{\mu - r}{\sigma^2}\left( C n_2(n_2-1) Y_t^{n_2-1} - \frac{2\bar{U}_1(Y_t)}{\theta^2 Y_t} \right),
		\end{align*}
		and
		\begin{align*}
			\tau^* = \inf\big\{t>0 :~ X_t \geq {x^*}\big\}.
		\end{align*}
	\end{theorem}
	\noindent Proof: see Appendix \ref{Pf_theorem_strategy}.
	
	\noindent In applications the optimal strategies $\left(c^*_t, b^*_t, \pi^*_t\right)$ can be determined using the mapping function given in (\ref{YtoX}) and the critical wealth level $x^*$ can be obtained from the numerical solution to $y^*$ using (\ref{ystar}) and the marginal utility (\ref{xstar}). We will present the numerical application of these results in the next section.
	
	\section{Implementation and numerical results}
	In this section, we demonstrate the analytic properties and numerical results of the presented model and results. In the first part, we present the value function and the optimal annuity under different labor and market environments by varying the pension scheme, labor rate, interest rate, and risk profiles. In the second part, we investigate model performance using simulated paths of the risky asset price. We choose a typical market scenario for investment and analyze the optimal strategies for different retirees.
	
	\subsection{Analytic results}
	We focus on the dependency of the critical level ${x^*}$ and the utility level $V(x)$ on all other parameters in our model. We have the utility preference $(\alpha,\beta,p_1,p_2)$, market profile $(r,\mu\,\sigma)$, pension scheme $(k,\delta)$, and labor choice $(w,\bar b)$ in our model. In addition, we introduce two weights $(v_1,v_2)$ to leverage the running utility and the stopping utility. A similar argument can be found in \citet{gerrard2012choosing}, however, we denote $v_2/v_1$ as the leverage ratio which will be explained later in this section.  In the following applications, we will apply the theoretical results on the leveraged value function (\ref{leveraged_value_function}). 
	\begin{equation}\label{leveraged_value_function}
		V(x) = \max_{c_t,b_t,\pi_t,\tau}\mathbb{E}^x\Bigg[v_1\int_0^{\tau}e^{-\rho t}U_1(c_t,b_t)dt + v_2\frac{e^{-\rho \tau}}{\rho}U_2(kX_\tau^{0,x})\Bigg].
	\end{equation}
	The common setting is for a male retiree at age 60 with mortality rate $\delta= 0.01$ and, unless otherwise specified, parameters:
	$$r = 0.035,~\mu = 0.08,~\sigma = 0.15,~\rho = 0.045,~k = 0.095,~w=100,~\bar b=0.5,~\alpha=0.5,~p_1 = p_2 = 2,~v_1=0.01,~v_2=0.1.$$
	The range for  various parameter combinations are $r\in[0.01,0.05]$,  $\mu\in[0.07,0.12]$, $\sigma\in[0.1,0.25]$, $\bar b\in[0,1]$, $\delta\in[0.01,0.022]$, $k\in[0.07,0.1]$, and $v_1\in[0.01,0.1]$. %

\begin{figure}[H]
		\centering
		\caption{Value Function with different $\beta$ and $\alpha$}
		\includegraphics[width=0.8\linewidth]{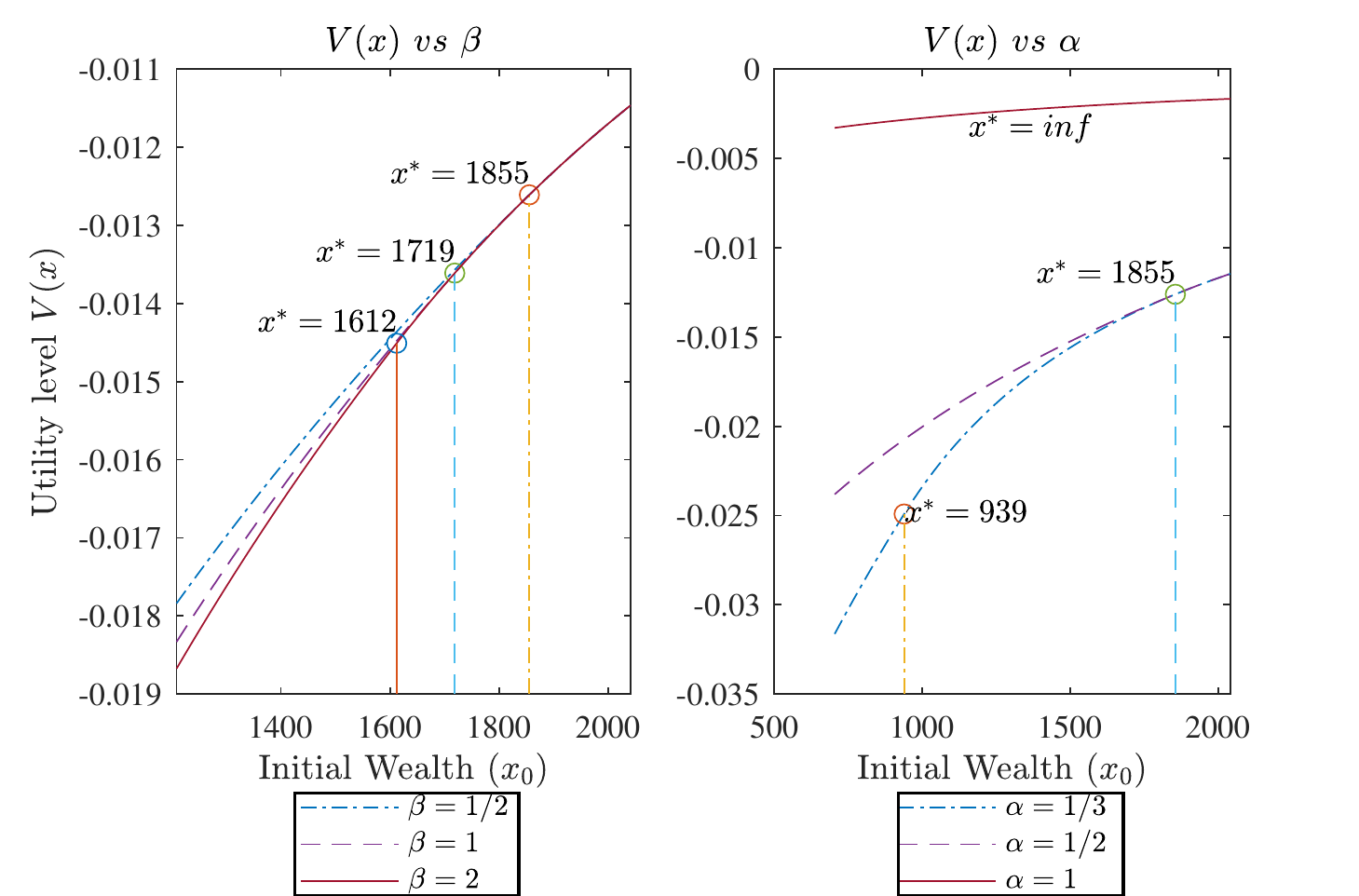}
		\label{chap1:pic1} %
	\end{figure}

\begin{figure}[H]
		\centering
		\caption{Value Function with different wage and labor rates}
		\includegraphics[width=0.8\linewidth]{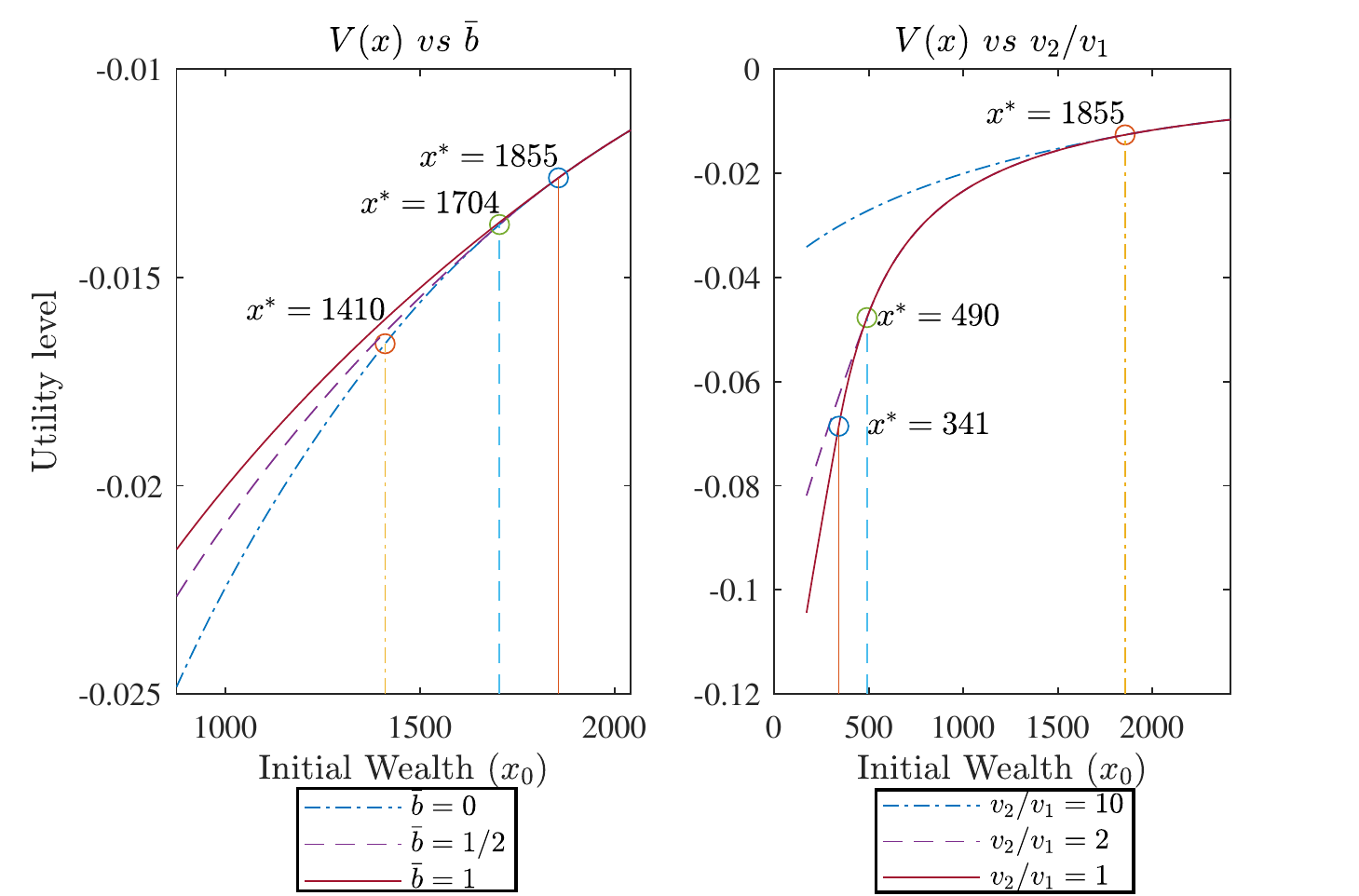}
		\label{chap1:pic2} %
	\end{figure}

\begin{figure}[H]
		\centering
		\caption{Annuity as function of interest rate and Sharpe ratio}
		\includegraphics[width=0.8\linewidth]{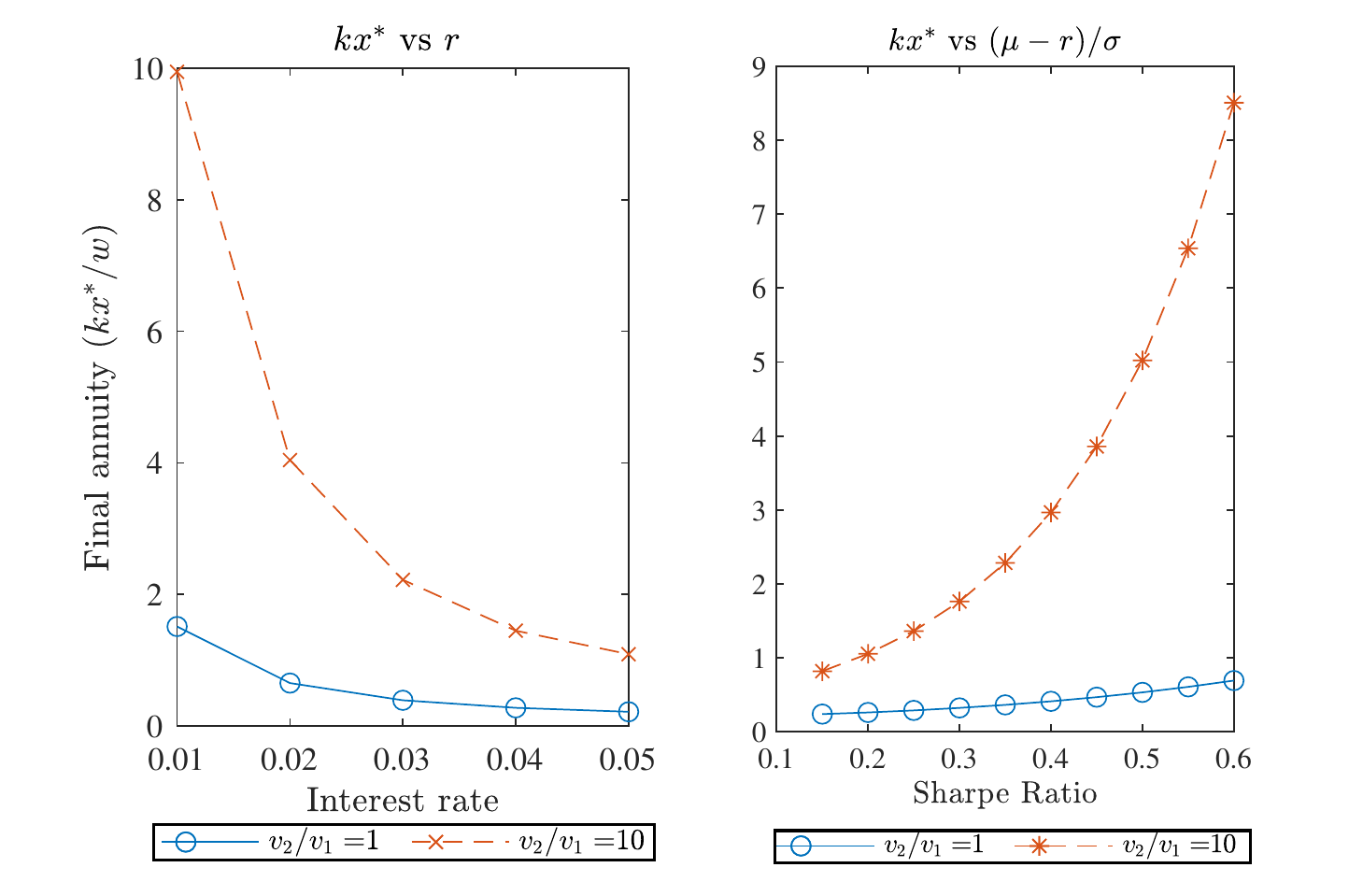}
		\label{chap1:pic3}
	\end{figure}

\begin{figure}[H]
		\centering
               	\caption{Critical level as function of mortality rate and pension scheme}
		\includegraphics[width=0.8\linewidth]{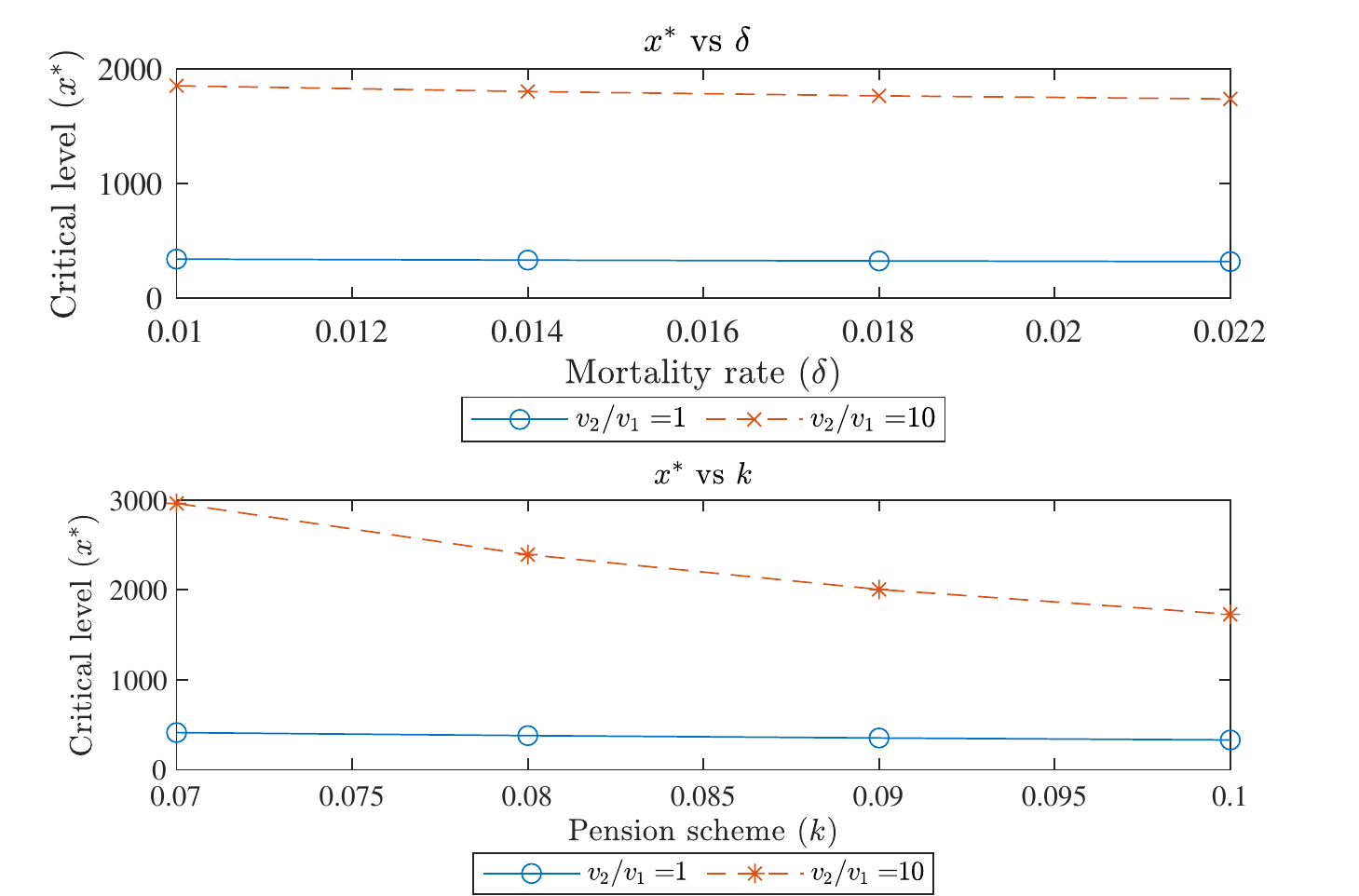}
		\label{chap1:pic4}
	\end{figure}

\begin{figure}[H]
		\centering
		\caption{Optimal annuity surface as a function of wage rate and labor rate}
		\includegraphics[width=0.8\linewidth]{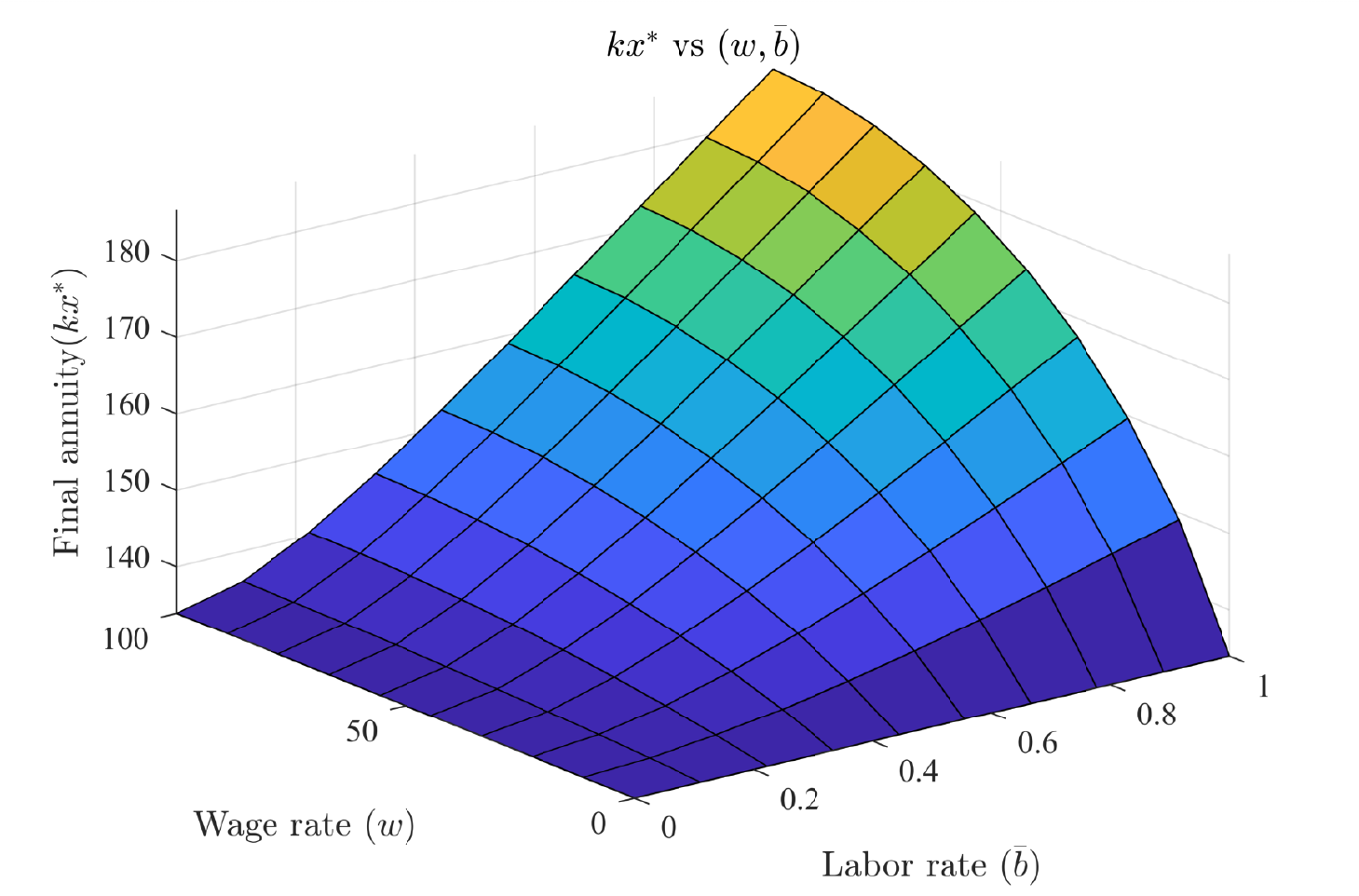}
		\label{chap1:pic5}
	\end{figure}

	Recalling the main theoretical results of Section~\ref{sec:3} we can make two immediate observations. First, by varying the value of $(w,\beta,\bar b)$, we can switch our model to include or exclude the labor income.  Second, by varying the value of $(\alpha,p_1,p_2)$, we can obtain solutions to the ruined cases where the retirees continue their optimal consumption, labor, and investing strategies without obtaining a stopping annuity. By the definition of the utility function, the parameter $\beta$ is the retiree's  labor preference and $\alpha$ is consumption preference parameter. The impact of the parameters are shown in Figures \ref{chap1:pic1}, \ref{chap1:pic2}, \ref{chap1:pic3} and \ref{chap1:pic4} and we make the following observations:
	\begin{itemize}
		\item [1)] Increasing the value of $\beta$, the retiree favors lower labor and the final annuity decreases, as shown in the left graph of Figure \ref{chap1:pic1}, the critical level decreases from 1855 to 1612 as $\beta$ increases from 0.5 to 2;
		\item[2)] Increasing the value of $\alpha$, the retiree favors higher consumption and the final annuity increases, as shown in the right graph of Figure \ref{chap1:pic1}, the critical level increases from 939 to 1855 as $\alpha$ increases from 1/3 to 1/2; continuing to increase the value of $\alpha$ will lead to ruined problem, when $\alpha=1$, the solid line in the right graph of Figure \ref{chap1:pic1} shows the ruined value function since the running utility will dominate the target function such that no stopping annuity applies;
		\item[3)] Increasing the maximum labor rate $\bar b$, the final annuity increases, as shown in the left graph of Figure \ref{chap1:pic2};
		\item[4)] Increasing the leverage ratio $v_2/v_1$, the final annuity increases substantially, as shown in the right graph of Figure \ref{chap1:pic2};
		\item[5)] Increasing the interest rate, the final annuity decreases, as shown in the left graph of Figure \ref{chap1:pic3}. Comparing the two curves in Figure \ref{chap1:pic3}, when the leverage ratio $v_2/v_1$ increases from 1 to 10, the effects of the interest rate and the Sharpe ratio are greatly amplified. Similar results hold in Figure \ref{chap1:pic4} for varying mortality rate $\delta$ and the pension scheme $k$.
		\item[6)] Increasing the Sharpe ratio the final annuity increases, as shown in the right graph of Figure \ref{chap1:pic3};
		\item[7)] Increasing the mortality rate $\delta$ or the pension scheme $k$, the critical level decreases, as shown in Figure \ref{chap1:pic4}. The almost flat curves in the top graph of Figure \ref{chap1:pic4} reflects that the mortality rate has very little effect in the model even with a higher leverage ratio;
		\item[8)] With the higher leverage ratio, the final annuity is concave increasing with respect to the wage rate and convex increasing with respect to the labor cap $\bar b$, as shown in Figure \ref{chap1:pic5}. However, with lower leverage ratio, the final annuity changes to concave increasing with respect to the labor cap $\bar b$. Due to limited space, we do not show this relationship here.
	\end{itemize}
	
	Figures~\ref{chap1:pic1}-\ref{chap1:pic5} illustrate some interesting facts about the model outputs. Utility preference $(\alpha,\beta,p_1,p_2)$ dominate the shape of the value function and determine the boundary of the optimal annuity where the stopping problem could become a ruined problem. The market profile $(r,\frac{\mu-r}{\sigma})$ has the opposite effect on the optimal annuity. As the interest rate increases, the retiree would prefer a lower optimal annuity and brings annuitization earlier. When the Sharpe ratio increases, the retiree would be likely to keep consuming and investing in the financial market and postpone annuitization to when their wealth reaches a higher value. Higher values of the leverage ratio $v_2/v_1$ amplify the effectiveness from the market profile, but has a small impact on the pension scheme $(\delta,k)$. Although the mortality ratio $\delta$ is a varies as the retiree ages, it does not substantially influence the choice of annuitization. For the labor choice $(w,\bar b)$ after retirement, the optimal annuity increases by over $30\%$ from $\bar b=0$ to $\bar b=1$ at wage rate $w=100$. However, the change to the optimal annuitization time is not clear, and this will be investigated in the following simulation study.

	\subsection{Simulation results}
	In this simulation study we set the leverage ratio at $v_2/v_1=10$ to induce a higher final annuity and fix the other parameters at a moderate level:
	$$r = 0.035,~\mu = 0.08,~\sigma = 0.15,~\rho = 0.045,~k = 0.095,~w=100,~\alpha=0.5,~p_1 = 2,~p_2 = 2,~v_1=0.01,~v_2=0.1.$$
	We compare our model under three different labor choices: 
	\begin{itemize}
		\item model 1 (denoted m1): no labor income $\bar b=0$ and critical level $x_1^*=1409.93$;
		\item model 2 (denoted m2): lower labor limit $\bar b=0.25$, lower labor preference $\beta=1$, and critical level $x_2^*=1613.22$; and
		\item model 3 (denoted m3): higher labor limit $\bar b=0.5$, higher labor preference $\beta=0.5$, and critical level $x_3^*=1855.29$.
	\end{itemize}
There is an approximately $15\%$ difference in critical levels $x^{*}$ from model 1 to model 2 and from model 2 to model 3.
	
\begin{figure}[H]
		\centering
		\caption{Expected annuitization time, annual consumption and labor income vs initial wealth}
		\includegraphics[width=0.8\linewidth]{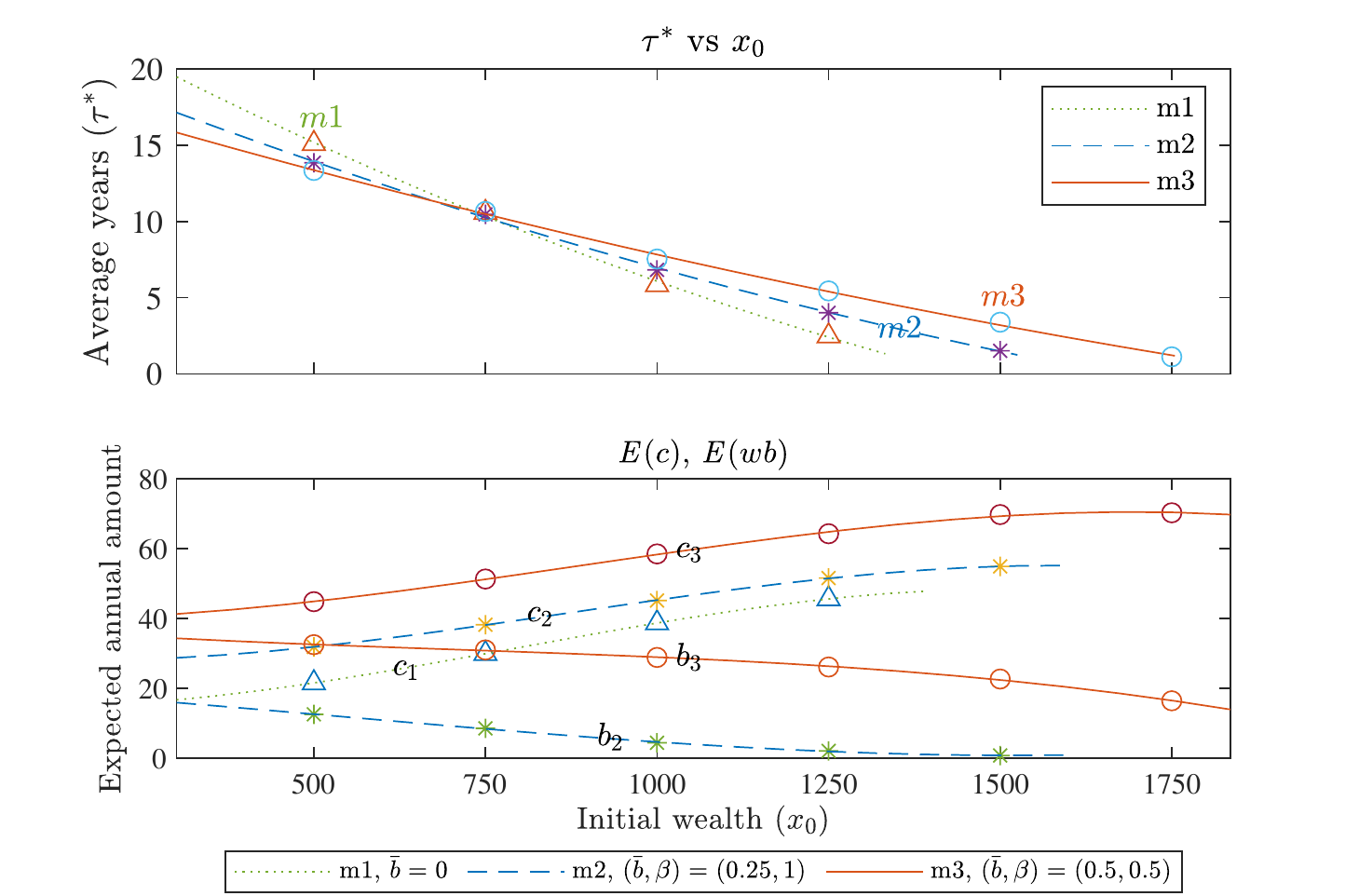}
		\label{chap1:pic6}
	\end{figure}

	We run 1000 Monte-Carlo simulations of the path of the risky asset as scenarios for.   We assume the retirees are males at age 60  and they follow the optimal strategies. To obtain accurate optimal annuitization times $\tau^*$, we run the simulation for up to 20 years. However, in other scenarios we force retirees to annuitize at 15 years. Initial wealth is chosen from the set $x_0\in\{500, 750, 1000, 1250, 1500, 1750\}$ and we run the simulations for each initial wealth. The optimal annuitization time $\tau^*$ is reported in in the top panel of Figure \ref{chap1:pic6}. We find that the optimal annuitization time $\tau^*$ seems to be linear to the initial wealth $x_0$ and the effect of the labor income is twofold. First, more labor income lifts the level of the optimal annuity. Second, labor income flattens the slope of the optimal annuitization time. As a result, retirees with lower initial wealth could benefit from earlier annuitization and a higher annuity payment. The lower panel in Figure \ref{chap1:pic6} presents the expected annual amount of consumption and income from labor. The expected annual consumption is linearly increasing with respect to the initial wealth and the expected annual income is linearly decreasing with respect to the initial wealth. Higher labor limits (high $\bar b$) and higher labor preferences (lower $\beta$) increase both the consumption rate and income rate.
	
\begin{figure}[H]
		\centering
		\caption{Average consumption and labor income over age}
		\includegraphics[width=0.8\linewidth]{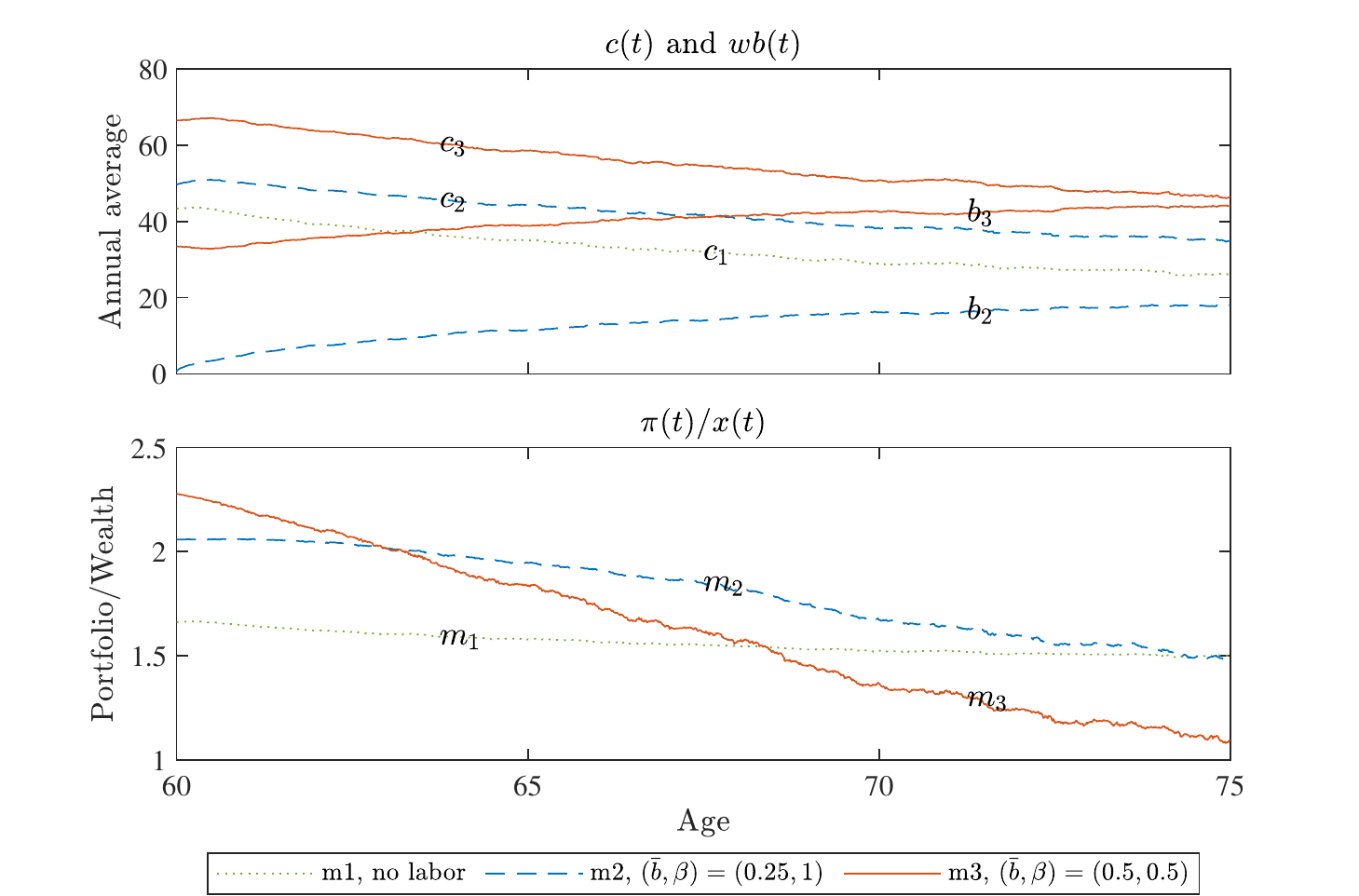}
		\label{chap1:pic7}
	\end{figure}

	From the six initial wealth levels shown in Figure \ref{chap1:pic6}, we fix $x_0=1000$ as the common initial wealth for the remaining analysis the three models. Since the optimal annuitization time of three models at $x_0=1000$ are within a moderate range between about 6 years and 8 years, the small differences in the optimal annuitization times will make the following comparisons simpler.  Figure \ref{chap1:pic7} reports the average consumption, income, and portfolios among 1000 retirees at different ages from 60 to 75. We find the retirees' behavior in consumption and labor are very similar over time. The amount of  consumption and the portfoliotends to decrease over time.  Labor income increases slightly at earlier times after retirement and flattens. The big difference between the three models lies in the investment strategies. Retirees who chose model 2 and model 3 experience a decline in their portfolio, and the decline is more significant in model 3. This decline in the portfolio could be explained by the labor income. Recall that the average annuitization time is around 6 years, so after age 65 the retiree whose wealth is still below the optimal annuity level needs to work more and reduce both consumption and investment at the same time in model 2 and model 3. Retirees who choose model 1 have no labor income so their portfolio remains relatively flat for the entire time period. Comparing the curves of model 2 and model 3 in the top panel of Figure \ref{chap1:pic7}, we find that the labor income in model 3 could compensate for most of the consumption at later period after retirement, this explains why the portfolio in model 3 declines faster than model 2.

	\begin{figure}[H]
		\centering
		\caption{Histogram: The distribution of the optimal annuitization time}
		\includegraphics[width=0.8\linewidth]{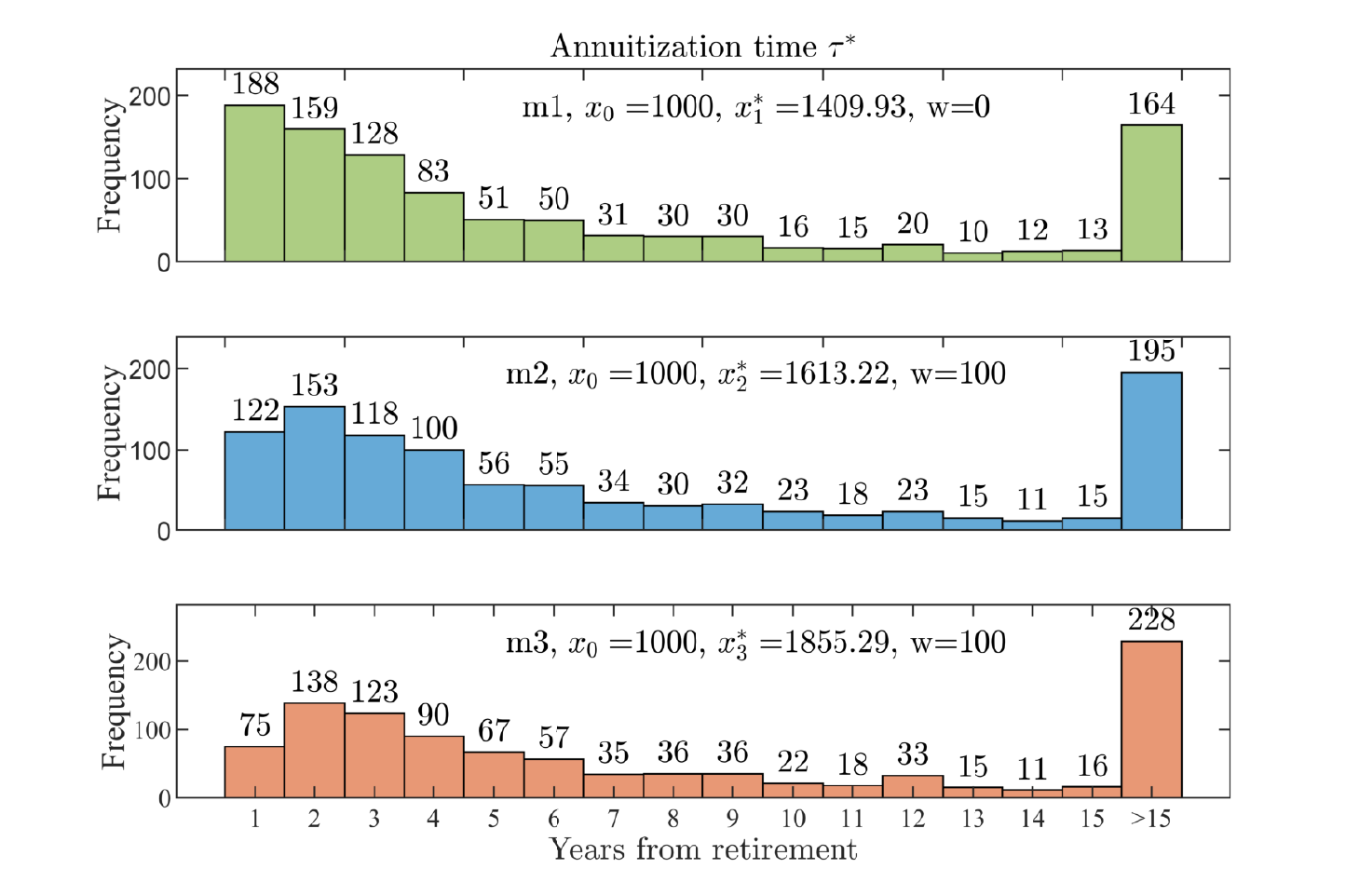}
		\label{chap1:pic8}
	\end{figure}
	
	\begin{figure}[H]
		\centering
		\vspace{-0.5cm}
		\caption{Distribution of the present value of total annuity}
		\includegraphics[width=0.8\linewidth]{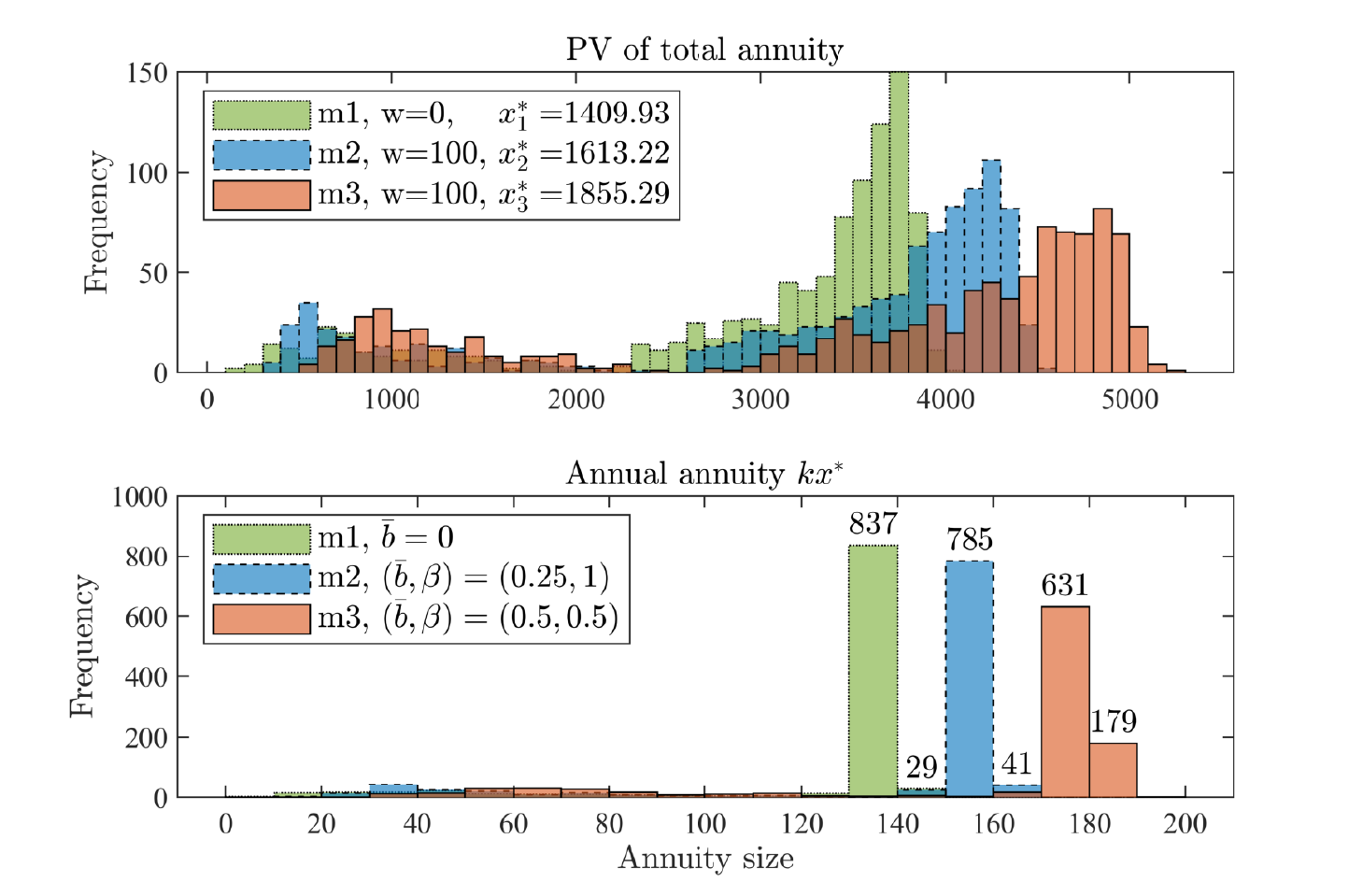}
		\label{chap1:pic9}
	\end{figure}
	
	Figure \ref{chap1:pic8} reports the distribution of the annuitization time for three models at initial wealth $x_0=1000$. The last column in each graph shows the number of retirees who failed to annuitize within 15 years. From those histograms, we compare the estimated probability of annuitization within 15 years and within 6 years of retirement. The empirical probabilities of annuitization in model 1 within 15 years and 6 years of retirement are $83.6\%$ and $65.9\%$, respectively, while the target annuity is $40\%$ higher than the initial wealth. The corresponding empirical probabilities in model 2 become $80.5\%$ and $60.4\%$, respectively, while the target annuity is increased by $14.5\%$ from that in model 1. The corresponding empirical probabilities of annuitization in model 3 are $77.2\%$ and $55\%$, respectively, while the target annuity is increased by $15\%$ from that in model 2.
	
	Figure \ref{chap1:pic9} reports the simulated distribution of annuities under two different measures. The top panel of Figure \ref{chap1:pic9} shows the frequencies of the present value of the total annuity and the bottom panel shows the frequencies of annual payment ampounts from annuitization for each model.  About $86.6\%$ retirees who choose model 1 receive an annual annuity payment of more than 130. About $82.6\%$ retirees who choose model 2 receive an annual annuity payment over 150. About $81\%$ retirees who choose model 3 receive an annual annuity payment greater than 170. Since we force retirees to annuitize after 15 years, there are two peaks for each model shown in the histograms in the top panel of Figure \ref{chap1:pic9}. The distributions from from each of these three models appear in shape so we use model 3 to illustrate the effects. The histogram of model 3 shows that the mean present value of the total annuity is around 4500 which relects the $81\%$ of retirees who receive an annual annuity payment over 170 and who take on average less than 6 years to annuitize. The remaining $29\%$ retirees whose total annuity lies between 0 to 4000 form a bimodal distribution with two peaks, the first located around 1000 and the second located around 3500.  The left tail of the distribution below 1000  reflects the annuities of retirees who are forced to annuitize at 75 ages and the minimum of the distribution to the left of the peak at 3500 reflects those retirees who take between between 6 to 15 years to annuitize.  From Figure \ref{chap1:pic9} we can observe that model 3 performs better than model 1 from the following perspectives: first, the median of the total annuity has increased from 3500 to 4500; second, the left tail leads to a higher average annuity in model 3 when annuitization fails; third, the peak of the distribution are around the median is reduced and the frequency of the total annuities are concentrated to the right of the median.
	
	\begin{table}[H]
		\centering
		\caption{Comparison of the expected present value of labor income, consumption, annuity and annuitization time}\label{chap1:table}
		\hspace*{-0.0cm}
		\begin{adjustbox}{width=0.9\linewidth,center}
			\begin{tabular}{|c|ccc|ccc|ccc|ccc|}
				\hline
				\multirow{2}{*}&\multicolumn{3}{|c|}{Labor income (annual)}&\multicolumn{3}{c|}{Consumption (annual)}&\multicolumn{3}{c|}{Annuity (annual)}&\multicolumn{3}{c|}{Net wealth (present)}\\
				\cline{2-13}
				&\multicolumn{1}{|c}{m1}& m2& m3 &m1& m2& m3&m1& m2& m3&m1& m2& m3\\
				\hline
				\hline
				min & 0 & 0.00 & 19.81 & 12.0 & 16.2 & 20.0 & 6.4 & 10.8 & 20.0 & 165 & 286 & 558 \\
				5th & 0.00 & 0.02 & 22.83 & 21.0 & 27.1 & 33.9 & 21.9 & 19.7 & 30.8 & 601 & 554 & 869 \\
				25th & 0.00 & 0.48 & 25.86 & 32.3 & 38.1 & 48.9 & 98.1 & 104.2 & 111.1 & 2772 & 2946 & 3154 \\
				50th & 0.00 & 3.05 & 29.50 & 40.5 & 46.9 & 60.9 & 120.6 & 134.3 & 148.6 & 3407 & 3793 & 4212 \\
				75th & 0.00 & 8.02 & 32.93 & 45.3 & 52.3 & 68.9 & 129.2 & 145.9 & 164.4 & 3645 & 4119 & 4658 \\
				90th & 0.00 & 13.29 & 35.94 & 47.0 & 55.2 & 72.5 & 132.7 & 150.8 & 171.2 & 3746 & 4255 & 4843 \\
				max & 0.00 & 18.58 & 39.69 & 51.0 & 59.1 & 77.5 & 141.9 & 159.2 & 182.1 & 4006 & 4492 & 5154 \\
				\hline
				\hline
				median & 0.00 & 3.05 & 29.50 & 40.5 & 46.9 & 60.9 & 120.6 & 134.3 & 148.6 & 3407 & 3793 & 4212 \\
				mean & 0.00 & 4.88 & 29.55 & 38.1 & 44.8 & 58.1 & 105.4 & 114.9 & 127.5 & 2972 & 3242 & 3615 \\
				std & 0.00 & 5.16 & 4.48 & 8.7 & 9.3 & 12.7 & 35.4 & 43.6 & 49.4 & 1003 & 1233 & 1398 \\
				
				\hline
			\end{tabular}
		\end{adjustbox}
	\end{table}

	Finally, we summarize the quantiles and other statistics from the three models in Table \ref{chap1:table}, which reports the present value of annual labor income, annual consumption, annual annuity payment, and the net wealth. Before annuitization, the retiree has labor income as an input cash flow and consumption as an output cash flow. After annuitization, the retiree receives the annuity payments as an input cash flow. The net wealth is calculated by subtracting the consumption from the labor income and annuity payments. Recall that a retiree who choose model 3 has a higher labor limit and higher preference to work.  Table \ref{chap1:table} shows that  model 3 performs better than model 1 and model 2 in every quantile. The higher cash flow from labor income allows the retiree who's behavior is reflected by model 3 to consume more and annuitize at a higher level than retirees represented by the labor attributes of model 1 and model 2.

	\section{Conclusion}
	In this paper, we study expected utility maximization and constrained post-retirement annuitization with extra labor income in the framework of Cobb-Douglas risk preferences. This is a constrained stochastic control and optimal stopping problem where the controls are  labor,  consumption, and  wealth invested in a risky asset and the stopping decision is the annuitization time. Using the martingale approac the dual problem is constructed as an unconstrained optimal stopping problem and is solved in closed form. The solution of the dual problem yields the solution to our constrained stochastic control/optimal stopping problem.
	
	In numerical applications, we investigate the effectiveness of the utility preference, market profile, pension scheme, labor choice, and leverage ratio to the model output. We find that a market with higher interest rates would decrease the willingness of retirees to continue consuming and working, and as such they are more likely to choose to annuitize earlier at a lower position. The leverage ratio is a useful tool to numerically analyze  model outputs, as a high leverage ratio could amplify the effect from a small change of model parameters. We find that our model is very insensitive to the mortality rate, as the optimal annuity changes very little, so we may ignore mortality variations post-retirement. We examine the optimal annuity surface as a function of the wage rate and the labor rate and observe that the optimal annuity is concave in the wage rate and convex in the labor rate. However, the leverage ratio could change the convexity in the labor rate to concavity.
	
	We perform a Monte-Carlo simulation study of three model environments for retirees to choose from: no work, work less, and work more. Simulations results show that the optimal annuitization time is almost linear with respect to the initial wealth. Extra income from post-retirement labor increases the final annuity, reduces the annuitization time, and changes the investment strategy. We find that the labor income will be kept at a certain level when the amount of consumption and investment decrease over time. We provide different measures of the final annuity premium, which show that once the optimal annuity is increased by $15\%$, the probability of annuitization within a fixed number years is decreased by $5\%$. The simulated empirical distribution of the the present value of the total annuity  shows that labor income after retirement improves the financial situation of retirees in the worst cases and their wealth is more evenly distributed.

	\bibliographystyle{abbrvnat}
	\bibliography{mybib}

	\appendix
	
	\section{Appendix}
	This appendix provides proofs of the technical lemmas and theorems.
	\subsection{Proof of Lemma \ref{equality_lemma}}
	\begin{proof}\label{Pf_lemma_equality_lemma}
		We define the following continuous process $X_t$ and by Bayes's rule we have
		\begin{align*}
			X_t&=\frac{1}{H_t}\mathbb{E}_\mathbb{P} \left[\int_{t\wedge \tau}^\tau {H_s\left(c_s + wl_s\right) ds} + H_\tau \left(B+\frac{w}{r}\right)  \Bigg |\mathcal{F}_t\right] - \frac{w}{r} \\
			&=\mathbb{E}_{\mathbb{Q}}\left[\int_{t\wedge \tau}^\tau {e^{-r(s-t)}\left(c_s + wl_s\right) ds} + e^{-r(\tau-t)} \left(B+\frac{w}{r}\right)  \Bigg |\mathcal{F}_t\right] - \frac{w}{r} ,
		\end{align*}
		Obviously we have $X_0 = x$ and $X_\tau = B$ $a.s$.
		
		By Fatou's lemma, the process inside the expectation is non-negative for all positive $B$ and any stopping time $\tau$
		\begin{equation*}
			\int_t^\tau {e^{-r(s-t)}\left(c_s + wl_s\right) ds} + e^{-r(\tau-t)} \left(B+\frac{w}{r}\right)   \geq 0,~\forall t\in[0,T],
		\end{equation*}
		which gives
		\begin{equation*}
			X_t\geq - \frac{w}{r}.
		\end{equation*}
		Similarly to \citet{karatzas2000utility}, we define the following process
		\begin{equation}\label{M_t_1}
			M(t) = e^{-rt}\left(X(t) + \frac{w}{r}\right) + \int_0^t e^{-rs}\left(c(s) + w l(s)\right) ds.
		\end{equation}
		Then $M(t)$ in (\ref{M_t_1}) is a martingale under $\mathbb{Q}$
		\begin{align*}
			\mathbb{E}_{\mathbb{Q}} \left[M(t)\big|\mathcal{F}(s)\right] =& \mathbb{E}_{\mathbb{Q}} \left[\int_0^\tau e^{-rt}\left(c(t) + w l(t)\right)dt + e^{-r\tau}\left(B+\frac{w}{r}\right)\big| \mathcal{F}(s)\right]\\
			=&e^{-rs}\mathbb{E}_{\mathbb{Q}} \left[\int_s^\tau e^{-r(t-s)}\left(c(t) + w l(t)\right)dt  + e^{-r(\tau - s)}\left(B+\frac{w}{r}\right)\big| \mathcal{F}(s)\right]\\
			&+ \int_0^s e^{-rt}\left(c(t) + w l(t)\right)dt\\
			=&e^{-rs}\left(X(s) + \frac{w}{r}\right) + \int_0^s e^{-rt}\left(c(t) + w l(t)\right)dt
			= M(s).
		\end{align*}
		By the martingale representation theorem, there exists a progressively measurable process $\phi(t)$ such that $\int_0^T \|\phi_t\|^2dt <\infty$ with probability one and $M(t)$ can be written as
		\begin{equation}\label{M_t_2}
			M(t) = x + \frac{w}{r} + \int_0^t \phi_s d\bar{B}_s.
		\end{equation}
		Let $\pi_t = e^{rt}\phi_t /\sigma$, then we can verify using equations (\ref{Xtprocess}), (\ref{M_t_1}) and (\ref{M_t_2}) that $X_t = X^{\left(\pi\right)}_t$ a.e.
	\end{proof}
	
	\subsection{Proof of Lemma \ref{Vbar_prime}}
	\begin{proof}\label{Pf_lemma_Vbar_prime}
		The convexity of $\bar{U}_1$ and $\bar{U}_2$ gives
		\begin{equation*}
			\bar{U}_i(x) - \bar{U}_i(y) \geq \bar{U}'_i(y) (x-y),~~~\forall x,~y>0,\text{ for } i=1,2.
		\end{equation*}
		By the definition of $\bar{U}_1$ and $\bar{U}_2$ we obtain
		\begin{align*}
			\bar{U}'_1(y) &= I_c(y) + wI_l(y),\\
			\bar{U}'_2 (\rho y) &= I\left(\rho y\right),
		\end{align*}
		For any real number $h$ satisfying $|h|<<y$, we have 
		\begin{align*}
			&\bar{V}(x,y+h) - \bar{V}(x,y) \\
			\geq &h\mathbb{E}\left[\int_0^{\tau^*_y} H_t \bar{U}_1'(Y_t^y) + H_{\tau^*_y}\left(\bar{U}'_2\left(\rho Y_{\tau^*_y}^y\right) - \frac{w}{r}\right)\right]+h(x+\frac{w}{r})\\
			=& -h\mathbb{E}\left[\int_0^{\tau^*_y} { H_t\left(I_c(Y_t^y) + w I_l(Y_t^y)\right)dt} + H_{\tau^*_y} \left(I\left(\rho Y_{\tau^*_y}^y\right) + \frac{w}{r}\right)\right] +h(x+\frac{w}{r})\\
			= &h\left(x-\mathbb{X}_{\tau^*_y}(y)\right)\\
			=& 0,
		\end{align*}
		which leads to
		\begin{align*}
			\lim_{h\rightarrow0^-}\frac{\bar{V}(x,y+h) - \bar{V}(x,y)}{h} \leq 0 \leq \lim_{h\rightarrow0^+}\frac{\bar{V}(x,y+h) - \bar{V}(x,y)}{h} .
		\end{align*}
		If $\bar{V}(x,y)$ is differentiable at $y$, then we have 
		\begin{align*}
			\frac{\partial \bar{V}}{\partial y}(x,y) = \lim_{h\rightarrow0^-}\frac{\bar{V}(x,y+h) - \bar{V}(x,y)}{h} = 	\lim_{h\rightarrow0^+}\frac{\bar{V}(x,y+h) - \bar{V}(x,y)}{h} = 0.
		\end{align*}
	\end{proof}

	\subsection{Proof of Theorem \ref{equality_thm}}
	\begin{proof}\label{Pf_theorem_euqality_thm}
		If the value function $V(x)$ is obtainable, then there exists a stopping time $\tau^*\in\mathbf{S}$ such that
		\begin{equation*}
			V(x) = \sup_{\tau} V(x|\tau) = V(x|\tau^*).
		\end{equation*}
		By Lemma \ref{budgethold}, there also exists some strategy $\pi$ such that the budget constraint hold for all $x>0$
		\begin{equation}\label{eqn1}
			\mathbb{E}\left[\int_0^\tau {H_t\left(c_t + wl_t\right) dt} + H_\tau \left(X^{\pi} +\frac{w}{r}\right) \right] = x + \frac{w}{r},
		\end{equation}
		which give us the following equality for all $y>0$
		\begin{equation*}
			V(x|\tau) = \bar V(x,y|\tau).
		\end{equation*}
		So we have the following inequality for all $\tau$
		\begin{equation}\label{eqn2}
			V(x) \geq V(x|\tau)= \bar V(x,y|\tau),
		\end{equation}
		which also implies that
		\begin{equation}\label{eqn3}
			V(x) \geq \sup_{\tau} \bar V(x,y|\tau) \geq \bar V(x,y) \geq \inf_{y>0} \bar V(x,y).
		\end{equation}
		Considering that (\ref{ineq1}) gives
		\begin{equation}\label{eqn33}
			V(x) \leq \inf_{y>0} \bar V(x,y),
		\end{equation}
		so from (\ref{eqn3}) and (\ref{eqn33}) we can obtain
		\begin{equation}\label{eqn4}
			V(x) = \inf_{y>0} \bar V(x,y).
		\end{equation}
		Next, we show the dual problem is obtainable. The optimal controls could be obtained from the conjugate form once the budget constraint holds with  equality. We need only show that the stopping time $\tau^*$  that is optimal for $V(x)$ is also optimal for $\bar V(x,y)$. For all other $\hat \tau\in\mathbf{S}\setminus\tau^*$, the inequality in (\ref{eqn2}) would be strictly
		\begin{equation*}
			V(x) > V(x|\hat\tau) = \bar V(x,y|\hat\tau)\geq \inf_{y>0} \bar V(x,y|\hat\tau).
		\end{equation*}
		If the equality in (\ref{eqn4}) holds for some $\tau\in\mathbf{S}$, it must be $\tau^*$
		\begin{equation*}
			\inf_{y>0}\bar V(x,y) = \inf_{y>0} \bar V(x,y|\tau^*),
		\end{equation*}
		and we conclude that $\tau^*$ is also optimal for the dual problem. The same reasoning can be used to from the dual problem to the original problem.  Therefore, the two problems are equivalent.
	\end{proof}
	
	\subsection{Proof of Theorem \ref{Verification}}
	\begin{proof}\label{Pf_theorem_Verification}
		Assume that $\phi(y)$ is the solution to (\ref{Variational}) and $\phi'(y)$ is absolutely continuous. First, we show that $\bar{V}(x,y)=\phi(y) + y\left(x + \frac{w}{r}\right)$ is the solution to the dual problem (\ref{dual}). 
		We apply It\^{o}'s formula to $e^{-\rho t}\phi(Y_t)$ where $Y_t$ start from $y$ at $t=0$
		\begin{equation*}
			e^{-\rho t}\phi(Y_t) = \phi(y) + \int_0^{t} e^{-\rho s}\theta Y_s \frac{\partial \phi}{\partial y}(Y_s) dB_s + \int_0^{t} e^{-\rho s}\left(-\rho\phi(Y_s) + \mathcal{L}\phi(Y_s)\right)ds,
		\end{equation*}
		then, we obtain the following definitive equation for all $t$
		\begin{align}\label{Vbar_1}
			\phi(y) = \mathbb{E}\left[e^{-\rho t}\phi(Y_t)   + \int_0^{t} e^{-\rho s}\left(\rho\phi(Y_s) - \mathcal{L}\phi(Y_s)\right)ds\right].
		\end{align}	
		Equation (\ref{Variational}) implies that $\phi(Y_t) \geq \frac{1}{\rho}\bar{U}_2(\rho Y_t) - \frac{w}{r}Y_t$ and $\rho \phi(Y_t) - \mathcal{L}\phi(Y_t) \geq \bar{U}_1(Y_t)$ for all $Y_t$. Hence, for all stopping times $\tau$, we have
		\begin{equation*}
			\phi(y)\geq \mathbb{E}\left[e^{-\rho\tau}\left(\frac{1}{\rho}\bar{U}_2(\rho Y_\tau) - \frac{w}{r}Y_\tau\right) + \int_0^{\tau} e^{-\rho s}\bar{U}_1(Y_s^y)ds\right],
		\end{equation*}
		which gives
		\begin{equation*}
			\phi(y) \geq \bar{V}(x,y)-y\left( x + \frac{w}{r} \right).
		\end{equation*}
		Let $t = \tau^* = \inf \{t\geq 0: Y_t^y \leq {y^*}\}$ in (\ref{Vbar_1}) and notice that $\phi(Y_{\tau^*}) = \frac{1}{\rho}\bar{U}_2(\rho Y_{\tau^*}) - \frac{w}{r}Y_{\tau^*}$ and $\rho\phi(Y_s) - \mathcal{L}\phi(Y_s) = \bar{U}_1(Y_s^y)$, so we have
		\begin{align*}
			\phi(y) =& \mathbb{E}\left[e^{-\rho\tau^*}\left(\frac{1}{\rho}\bar{U}_2(\rho Y_{\tau^*}) - \frac{w}{r}Y_{\tau^*}\right)   + \int_0^{\tau^*} e^{-\rho s}\bar{U}_1(Y_s^y) ds\right]\\
			&\leq \bar{V}(y) - y\left( x + \frac{w}{r} \right).
		\end{align*}
		Therefore, $\bar{V}(x,y) = \phi(y) + y\left( x + \frac{w}{r} \right)$ and the equality holds at $\tau^*$.

	\end{proof}
	
	\subsection{Proof of Lemma \ref{existence}}
	\begin{proof}\label{Pf_lemma_existence}
		We denote
		\begin{equation*}
			\bar{U}(x) = \bar{U}_1(x) + \frac{n(p'_2)}{\rho}\bar{U}_2(\rho x) + wx,
		\end{equation*}
		and
		\begin{equation*}
			F(x) = \int_{+\infty}^{{x}}\frac{\bar{U}(z)}{z^{n_1+1}}(z)dz
		\end{equation*}
		If $0<p'_1<p'_2<1$, then there is a unique $y_0\in(0,\infty)$ such that $\bar{U}(y_0)=0$ and
		\begin{equation*}
			\left\{
			\begin{alignedat}{2}
				&\bar{U}(y)>0,~~~~~~~~&&y_0<y\\
				&\bar{U}(y)<0,~~~~~~~~&&0<y<y_0.
			\end{alignedat}
			\right.
		\end{equation*}
		
		Since
		\begin{equation*}
			\lim_{x\rightarrow 0^+}\frac{\bar{U}(x)}{x^{n_1+1}} =\lim_{x\rightarrow 0^+} x^{p'_1-(1+n_1)}\left(\tilde{A}-\frac{n(p'_2)}{\rho p'_2 k^{p'_2}}x^{p'_2-p'_1}\right)= -\infty,
		\end{equation*}
		and
		\begin{equation*}
			\lim_{x\rightarrow +\infty}\frac{\bar{U}(x)}{x^{n_1+1}} = 0^+,
		\end{equation*}
		that implies
		\begin{equation*}
			\lim_{x\rightarrow 0^+} F(x) = +\infty,
		\end{equation*}
		and
		\begin{equation*}
			\lim_{x\rightarrow +\infty} F(x) = 0^-.
		\end{equation*}
		Thus, there exists a $y^*\in(0,y_0)$ such that $F(y^*)=0$.
		
		Suppose there exists another $y_*\in(0,y_0)$ such that $F(y_*)=0$, then we get
		\begin{equation*}
			\int_{y_*}^{y^*} \frac{\bar{U}(z)}{z^{n_1+1}}dz =0,
		\end{equation*}
		and since $\bar{U}$ is negative on $(0,y_0)$, we must have $y_*=y^*$. Therefore, the solution of $y^*$ is unique and
		\begin{equation*}
			\left\{
			\begin{alignedat}{2}
				&F(y)\geq0,~~&&y\leq y^*\\
				&F(y)\leq0,~~&&y\geq y^*.
			\end{alignedat}
			\right.
		\end{equation*}
	\end{proof}
	
	\subsection{Proof of Lemma \ref{uniqueness}}
	\begin{proof}\label{Pf_lemma_uniqueness}
		First, we show that the following ODE has a unique solution
		\begin{equation}\label{ODE1}
			-\rho \phi(y)  + (\rho - r)y \phi(y) ' + \frac{1}{2}\theta^2 y^2\phi(y) '' + \bar{U}_1(y) = 0.
		\end{equation}
		We first consider the general solution to the following ODE
		\begin{equation}\label{ode1}
			-\rho \phi(y)  + (\rho - r)y \phi(y) ' + \frac{1}{2}\theta^2 y^2\phi''(y) = 0.
		\end{equation}
		Solving (\ref{ode1}), we obtain
		\begin{equation}\label{part1}
			\phi(y) = C_1 y^{n_1} + C_2 y^{n_2},
		\end{equation}
		where $C_1$ and $C_2$ are some constant that is determined from boundary value ${y^*}$, $n_1>1$ and $n_2\leq -\frac{2\left(\rho - r\right)}{\theta^2}<0$ are the two roots of function $n(x)$
		\begin{equation}\label{n(x)}
			n(x) = \frac{1}{2}\theta^2 x^2 + (\rho - r - \frac{1}{2}\theta^2)x -\rho.
		\end{equation}
		A particular solution for (\ref{ODE1}) is given by
		\begin{equation}\label{part2}
			\phi(y) = \frac{2}{\theta^2(n_1 - n_2)}\left(y^{n_1}\int {\frac{-\bar{U}_1(y)}{y^{n_1 +1}}}dy - y^{n_2}\int {\frac{-\bar{U}_1(y)}{y^{n_2 +1}}}dy\right).
		\end{equation}
		Combining (\ref{part1}) and (\ref{part2}), we have the general solution to (\ref{ODE1}),
		\begin{equation}\label{sol1}
			\phi(y) = C_1 y^{n_1} + C_2 y^{n_2} 
			+  \frac{2}{\theta^2(n_1 - n_2)}\left(y^{n_1}\int_{{y^*}}^y {\frac{-\bar{U}_1(z)}{z^{n_1 +1}}}dz - y^{n_2}\int_{{y^*}}^y {\frac{-\bar{U}_1(z)}{z^{n_2 +1}}}dz\right).
		\end{equation}
		
		For the boundary problem we show that for some positive ${y^*}$ the solution given in (\ref{sol1}) satisfies the boundary condition $\phi(y)=\frac{1}{\rho}\bar{U}_2(\rho y) - \frac{w}{r}y$, for $0 < y \leq {y^*}$. By investigating the convexity of $\bar{V}$
		\begin{align*}
			\phi'(y) = &\left(C_1 n_1 +\frac{2n_1}{\theta^2(n_1 - n_2)}\int_{{y^*}}^y {\frac{-\bar{U}_1(z)}{z^{n_1 +1}}}dz \right)y^{n_1-1}\\
			+ &\left(C_2 n_2 - \frac{2n_2}{\theta^2(n_1 - n_2)}\int_{{y^*}}^y {\frac{-\bar{U}_1(z) }{z^{n_2 +1}}}dz\right)y^{n_2-1},
		\end{align*}
		we find that $\phi(y)$ is strictly convex and decreasing. Thus the first term containing $y^{n_1-1}$ in (\ref{sol1}) must be zero as $y\rightarrow +\infty$
		\begin{equation*}
			\lim_{y\rightarrow +\infty}C_1 n_1 +\frac{2n_1}{\theta^2(n_1 - n_2)}\int_{{y^*}}^y {\frac{-\bar{U}_1(z)}{z^{n_1 +1}}}dz = 0,
		\end{equation*}
		and we obtain
		\begin{equation}\label{C1}
			C_1 = \frac{2}{\theta^2(n_1-n_2)}\int_{+\infty}^{{y^*}} {\frac{-\bar{U}_1(z)}{z^{n_1 +1}}}dz.
		\end{equation}
		Using (\ref{C1}) we may simplify (\ref{sol1}) with some constant $C$ as
		\begin{equation}
			\phi(y) = C y^{n_2} 
			+  \frac{2y^{n_1}}{\theta^2(n_1 - n_2)}\int_{+\infty}^y {\frac{-\bar{U}_1(z)}{z^{n_1 +1}}}dz - \frac{2y^{n_2}}{\theta^2(n_1 - n_2)}\int_{{y^*}}^y {\frac{-\bar{U}_1(z)}{z^{n_2 +1}}}dz.
		\end{equation}
		Applying the smooth connected conditions at ${y^*}$, we obtain
		\begin{equation}\label{C}
			C = C\left(y^*\right)=
			\frac{{y^*}^{-n_2}}{n_1-n_2}\left((n_1-p'_2)\frac{1}{\rho}\bar{U}_2(\rho{y^*}) -(n_1-1) \frac{w}{r}{y^*}\right),
		\end{equation}
		and
		\begin{equation}\label{ystar2}
			\frac{2{y^*}^{n_1}}{\theta^2}\int_{+\infty}^{{y^*}}\frac{-\bar{U}_1(z)}{z^{n_1 +1}}dz= y^* U'_2\left(\rho {y^*}\right) - \frac{n_2}{\rho}\bar{U}_2(\rho{y^*}) - (1-n_2)\frac{w}{r}{y^*}.
		\end{equation}
		Since we can rewrite
		\begin{equation}\label{simple1}
			y^*U'_2(\rho y^*) - \frac{n_2}{\rho}\bar{U}_2(\rho{y^*}) = \frac{p'_2 - n_2}{\rho}\bar{U}_2(\rho{y^*}) = 	\frac{2{y^*}^{n_1}}{\theta^2}\int_{+\infty}^{{y^*}} \frac{\frac{n(p'_2)}{\rho}\bar{U}_2(\rho z)}{z^{n_1+1}}dz,
		\end{equation}
		and
		\begin{equation}\label{simple2}
			(1-n_2)\frac{w}{r}{y^*}=\frac{(1-n_2)(1-n_1)}{r}\frac{wy^*}{1-n_1}=\frac{2{y^*}^{n_1}}{\theta^2}\int_{+\infty}^{{y^*}}\frac{n_1\frac{w}{r}z}{z^{n_1 +1}}dz = -\frac{2{y^*}^{n_1}}{\theta^2}\int_{+\infty}^{{y^*}}\frac{wz}{z^{n_1 +1}}dz.
		\end{equation}
		We simplify (\ref{ystar2}) using (\ref{simple1}) and (\ref{simple2})
		\begin{equation}\label{simple_y}
			\frac{2{y^*}^{n_1}}{\theta^2}\int_{+\infty}^{{y^*}} {\frac{\bar{U}_1(z) + \frac{n(p'_2)}{\rho}\bar{U}_2(\rho z) + wz}{z^{n_1 +1}}}dz = 0.
		\end{equation}
		By Lemma \ref{existence}, there exists a unique $y^*>0$ that solves equation (\ref{simple_y}). Therefore, $\phi(y)$ with $C$ given in (\ref{C}) and the value $y^*$ determined by (\ref{simple_y}) satisfies the boundary condition.

		Second, we show that $\phi(y)$ and ${y^*}$ are the optimal solutions to the dual problem by showing that $\phi(y)$ and ${y^*}$ solve the variational inequalities. We denote
		\begin{equation*}
			\Phi(y) = \phi(y) - \left(\frac{1}{\rho}\bar{U}_2(\rho{y}) - \frac{w}{r}{y}\right).
		\end{equation*}
		For $0<y<y^*$, we have 
		\begin{align*}
			-\rho \phi(y) + \mathcal{L}\phi(y) + \bar{U}_1(y) &= -\rho\left(\frac{1}{\rho}\bar{U}_2(\rho{x}) + \frac{w}{r}{x}\right) + \mathcal{L}\left(\frac{1}{\rho}\bar{U}_2(\rho{x}) + \frac{w}{r}{x}\right) + \bar{U}_1(y)\notag\\
			&=\bar{U}_1(y) + \frac{n(p'_2)}{\rho}\bar{U}_2(\rho y) + wy\notag\\
			&=\bar{U}(y)\notag\\
			&\leq 0.
		\end{align*}
		For $y^*\leq y$, we have
		\begin{align}
			-\rho \Phi(y) + \mathcal{L}\Phi(y) &= -\rho\left(\phi(y) - \frac{1}{\rho}\bar{U}_2(\rho{y}) + \frac{w}{r}{y}\right) + \mathcal{L}\left(\phi(y) - \frac{1}{\rho}\bar{U}_2(\rho{y}) + \frac{w}{r}{y}\right)\notag\\
			&= -\rho\phi(y) + \mathcal{L}\phi(y) -\left( \frac{n(p'_2)}{\rho}\bar{U}_2(\rho y) + wy\right)\notag\\
			&=-\left( \bar{U}_1(y) + \frac{n(p'_2)}{\rho}\bar{U}_2(\rho y) + wy\right)\notag\\
			&=-\bar{U}(y).\label{thisode}
		\end{align}
		Since $\Phi(y^*)=0$ and $\Phi'(y^*)=0$, the (\ref{thisode}) implies
		\begin{equation*}
			\Phi(y) = \frac{2y^{n_1}}{\theta^2(n_1 - n_2)}\int_{y^*}^y {\frac{-\bar{U}(z)}{z^{n_1 +1}}}dz - \frac{2y^{n_2}}{\theta^2(n_1 - n_2)}\int_{{y^*}}^y {\frac{-\bar{U}(z)}{z^{n_2 +1}}}dz.
		\end{equation*}
		Denote
		\begin{equation}\label{eeee}
			\psi(y) = y^{-n_2} \Phi(y)=\frac{2y^{n_1-n_2}}{\theta^2(n_1 - n_2)}\int_{y^*}^y {\frac{-\bar{U}(z)}{z^{n_1 +1}}}dz - \frac{2}{\theta^2(n_1 - n_2)}\int_{{y^*}}^y {\frac{-\bar{U}(z)}{z^{n_2 +1}}}dz.
		\end{equation}
		Differentiating (\ref{eeee}), we obtain
		\begin{align*}
			\psi'(y) =&\frac{2y^{n_1-n_2-1}}{\theta^2}\int_{y^*}^y {\frac{-\bar{U}(z)}{z^{n_1 +1}}}dz -\frac{2y^{n_1-n_2}}{\theta^2(n_1 - n_2)} \cdot{\frac{\bar{U}(y)}{y^{n_1 +1}}} + \frac{2}{\theta^2(n_1 - n_2)} \cdot{\frac{\bar{U}(y)}{y^{n_2 +1}}}\\ =&\frac{2y^{n_1-n_2-1}}{\theta^2}\int_{y^*}^y {\frac{-\bar{U}(z)}{z^{n_1 +1}}}dz
			= -\frac{2y^{n_1-n_2-1}}{\theta^2}F(y).
		\end{align*}
		Since $F(y) \leq 0$ for $y\geq y^*$, we have
		\begin{equation*}
			\psi'(y)\geq 0,
		\end{equation*}
		which implies $\psi(y)$ is monotone increasing function on $[y^*,\infty]$.
		\begin{equation*}
			\Phi(y) = y^{n_2} \psi(y) \geq y^{n_2} \psi(y^*)=0.
		\end{equation*}
		Therefore, we have $\phi(x) \geq \left(\frac{1}{\rho}\bar{U}_2(\rho{x}) - \frac{w}{r}{x}\right)$ for $y\in[y^*,\infty)$.
	\end{proof}
	
	\subsection{Proof of Theorem \ref{strategy}}
	\begin{proof}\label{Pf_theorem_strategy}
		The optimal stopping time $\bar{\tau}$ is a straightforward result of Theorem 10.4.1 of \citet{oksendal2003stochastic}. The optimal consumption and labor income are direct results from the dual problem. So we only need to show the optimal portfolio processes is generated by the optimal wealth processes (\ref{YtoX}). Applying It\^{o}'s formula to the optimal wealth process (\ref{YtoX}), we obtain
		
		\begin{align}
			dX_t^* =& \left(-(\rho-r)Y_t\phi^{\prime \prime}\left(Y_{t}\right)- \frac{1}{2}\theta^2Y^2_t\phi^{\prime \prime\prime}\left(Y_{t}\right)\right)dt + \theta Y_t \phi^{\prime \prime}\left(Y_{t}\right) dB_t\notag\\
			=& -\Bigg(C n_2 \left((\rho-r)(n_2-1) +\frac{1}{2}\theta^2(n_2-1)(n_2-2) \right)Y_t^{n_2-1}\notag\\
			&+\frac{2n_1Y_t^{n_1-1}}{\theta^2(n_1 - n_2)}\left((\rho-r)(n_1-1) +\frac{1}{2}\theta^2(n_1-1)(n_1-2) \right)\int_{+\infty}^{Y_t} {\frac{-\bar{U}_1(z)}{z^{n_1 +1}}}dz\notag\\
			&-\frac{2n_2Y_t^{n_2-1}}{\theta^2(n_1 - n_2)}\left((\rho-r)(n_2-1) +\frac{1}{2}\theta^2(n_2-1)(n_2-2) \right)\int_{{y^*}}^{Y_t} {\frac{-\bar{U}_1(z)}{z^{n_2 +1}}}dz\notag\\
			&+\frac{2\bar{U}_1(Y_t)}{Y_t} - \bar{U}'_1(Y_t) \Bigg)dt + \theta Y_t \phi^{\prime \prime}\left(Y_{t}\right) dB_t\notag\\
			=&\left(rX\left(X_t^* + \frac{w}{r}\right) + \bar{U}'_1(Y_t)\right)dt + \theta Y_t \phi^{\prime \prime}\left(Y_{t}\right) dB_t\notag\\
			& -\Bigg(C n_2 \left((\rho-r)n_2 -\rho +\frac{1}{2}\theta^2(n_2-1)(n_2-2) \right)Y_t^{n_2-1}\notag\\
			&+\frac{2n_1Y_t^{n_1-1}}{\theta^2(n_1 - n_2)}\left((\rho-r)n_1 -\rho +\frac{1}{2}\theta^2(n_1-1)(n_1-2) \right)\int_{+\infty}^{Y_t} {\frac{-\bar{U}_1(z)}{z^{n_1 +1}}}dz\notag\\
			&-\frac{2n_2Y_t^{n_2-1}}{\theta^2(n_1 - n_2)}\left((\rho-r)n_2 -\rho +\frac{1}{2}\theta^2(n_2-1)(n_2-2) \right)\int_{{y^*}}^{Y_t} {\frac{-\bar{U}_1(z) }{z^{n_2 +1}}}dz\notag\\
			&+\frac{2\bar{U}_1(Y_t)}{Y_t} \Bigg)dt \notag\\
			=&(rX_t^* + w + \bar{U}'_1(Y_t))dt + \theta Y_t \phi^{\prime \prime}\left(Y_{t}\right) dB_t\notag\\
			& -\Bigg(C n_2 \left(-\frac{1}{2}\theta^2n_2(n_2-1) +\frac{1}{2}\theta^2(n_2-1)(n_2-2) \right)Y_t^{n_2-1}\notag\\
			&+\frac{2n_1Y_t^{n_1-1}}{\theta^2(n_1 - n_2)}\left(-\frac{1}{2}\theta^2n_1(n_1-1) +\frac{1}{2}\theta^2(n_1-1)(n_1-2) \right)\int_{+\infty}^{Y_t} {\frac{-\bar{U}_1(z)}{z^{n_1 +1}}}dz\notag\\
			&-\frac{2n_2Y_t^{n_2-1}}{\theta^2(n_1 - n_2)}\left(-\frac{1}{2}\theta^2n_2(n_2-1) +\frac{1}{2}\theta^2(n_2-1)(n_2-2) \right)\int_{{y^*}}^{Y_t} {\frac{-\bar{U}_1(z)}{z^{n_2 +1}}}dz\notag\\
			&+\frac{2\bar{U}_1(Y_t)}{Y_t} \Bigg)dt\notag\\
			=&(rX_t^* + w +\bar{U}'_1(Y_t))dt + \theta Y_t \phi^{\prime \prime}\left(Y_{t}\right) dB_t\notag\\
			&+\theta^2 Y_t\Bigg( C n_2 (n_2-1)Y_t^{n_2-2}-\frac{2\bar{U}_1(Y_t)}{\theta^2Y_t^2}\notag\\
			&+\frac{2n_1(n_1-1)Y_t^{n_1-2}}{\theta^2(n_1 - n_2)}\int_{+\infty}^{Y_t} {\frac{-\bar{U}_1(z)}{z^{n_1 +1}}}dz - \frac{2n_2(n_2-1)Y_t^{n_2-2}}{\theta^2(n_1 - n_2)} \int_{{y^*}}^{Y_t} {\frac{-\bar{U}_1(z)}{z^{n_2 +1}}}dz \Bigg)dt \notag\\
			=&\left(rX_t^* + ( \mu  - r)\frac{\theta}{\sigma} Y_t \phi^{\prime \prime}\left(Y_{t}\right)  + \bar{U}'_1(Y_t) + w\right)dt + \theta Y_t \phi^{\prime \prime}\left(Y_{t}\right) dB_t.\label{aaaa}
		\end{align}
		Using the fact that
		\begin{equation*}
			\bar{U}_1(Y_t)' = -c_t^* -l_t^*w,
		\end{equation*}
		we obtain from (\ref{aaaa}) 
		\begin{equation*}
			dX_t^* = \left(rX_t^* + ( \mu  - r)\pi^* - c^* + \left(1-l^*\right)w \right)dt + \sigma\pi^* dB_t,
		\end{equation*}
		and
		\begin{align*}
			\pi_t^* =& \frac{\theta}{\sigma} Y_{t} \phi^{\prime\prime}\left(Y_{t}\right)\\
					=& \frac{\theta}{\sigma}\Bigg( C n_2 (n_2-1)Y_t^{n_2-1}-\frac{2\bar{U}_1(Y_t)}{\theta^2Y_t}\\
					&+\frac{2n_1(n_1-1)Y_t^{n_1-1}}{\theta^2(n_1 - n_2)}\int_{+\infty}^{Y_t} {\frac{-\bar{U}_1(z)}{z^{n_1 +1}}}dz-\frac{2n_2(n_2-1)Y_t^{n_2-1}}{\theta^2(n_1 - n_2)}\int_{{y^*}}^{Y_t} {\frac{-\bar{U}_1(z) }{z^{n_2 +1}}}dz \Bigg).
		\end{align*}
		The optimal wealth process and optimal controls are verified.
	\end{proof}
	
\end{document}